# A PRIMER FOR BLOCKCHAIN

Lawrence J. Trautman*
Mason J. Molesky**

## I. OVERVIEW

Much has been written about the likely impact of blockchain technology during its brief, decade-long existence. Early scholarship on this topic has focused on the legal and financial implications of virtual currencies which is based on blockchain technology. Rapid introduction and diffusion of technological changes throughout society, such as the blockchain, continue to exceed the ability of law and regulation to keep pace. Will blockchain prove as disruptive to business models and entrenched societal institutions as: electricity, radio, television, or the Internet? What beneficial aspects of the blockchain have been identified thus far and what future applications are probable? Should we expect blockchain technology to result in massive global changes? The primary goal of this article about blockchain is to present an all-encompassing basic explanation of what it is; how it works; why it's important; and identified potential uses to date.

Our article proceeds in six sections. First, we provide a summary description of the blockchain. Second is a discussion of the mechanics of this disruptive technology. Third, we present an analysis of two distributed ledger technologies, bitcoin and the IOTA protocol. Fourth is coverage of the history and use of virtual currencies. Fifth, we look at internet scams and the regulators charged with protecting against them. Sixth is a brief overview of the many identified uses to date for blockchain technology. And last, we conclude. We believe this paper is a valuable addition to the literature by providing a readable introduction and overview of what is now known about the likely impact of blockchain technology.

## II. WHAT IS THE BLOCKCHAIN?

Aaron Wright and Primavera De Filippi write, "[t]he blockchain is a distributed, shared, encrypted-database that serves as an irreversible and incorruptible public repository of information. It enables, for the first time, unrelated people to reach consensus on the occurrence of a particular transaction or event without the need for a controlling authority."[1] In essence, blockchain is

* BA, The American University; M.B.A., The George Washington University; J.D., Oklahoma City University School of Law. Mr. Trautman is an author and educator. He may be contacted at Lawrence.J.Trautman@gmail.com.
** B.S. (mathematics and computer science), Alma College; M.S. (cybersecurity), Ph.D. Candidate (computer science), The George Washington University. He may be contacted at masonmolesky@gmail.com.

The authors wish to express their thanks to the Research Roundtable on the Future of Financial Regulation held at George Mason University Antonin Scalia Law School, February 28 – March 1, 2019 and the following participants who provided inspiration for this article: Matthew Bruckner, Chris Brummer, A. Patrick Doyle, M. Todd Henderson, Julie Hill, Sarah Jane Hughes, Jamil N.

simply a data structure that leverages hash functions and encryption to provide the security of information like never seen before. Valentina Gatteschi notes the progression of blockchain technology:

> Three different blockchain evolutions can be identified: Blockchain 1.0, 2.0, and 3.0. Blockchain. . . . Blockchain 2.0 is about registering, confirming, and transferring contracts or properties. Application fields range from the use of blockchain as a decentralized copy of local databases (especially for public records and attestations) to more sophisticated applications. The most relevant feature of Blockchain 2.0 is the integration with smart contracts 1.0 is strongly related to Bitcoin and cryptocurrencies. . . . In Blockchain 3.0, the application field is no longer restricted to finance and goods transactions, but embraces sectors like government, health, science, education, and more."[2]

Gatteschi provides the following major advantages and disadvantages of blockchain:

---

Jaffer, Andrew Kloster, Robert H. Ledig, Soo Lee, Richard B. Levin, William J. Magnuson, Richard Neiman, Jeremy Newell, Saule T. Omarova, Elizabeth Rosenberg, Jeffrey Smith, Timothy Spangler, Thomas P. Vartanian, Angela C. Walch, Larry D. Wall, and Peter Wayner.

*** Much has been written about the likely impact of blockchain technology. Early scholarship on this topic has focused on the legal and financial implications of virtual currencies which is based on blockchain technology. Rapid introduction and diffusion of technological changes throughout society, such as the blockchain, continue to exceed the ability of law and regulation to keep pace. Will blockchain prove as disruptive to business models and entrenched societal institutions as: electricity, radio, television, or the Internet? What beneficial aspects of the blockchain have been identified thus far and what future applications are probable? Should we expect blockchain technology to result in massive global changes? The primary goal of this article about blockchain is to present an all-encompassing basic explanation of what it is; how it works; and why it's important. This article contributes to the scholarly literature and our understanding by providing a cogent description of this important technology and look at potential uses.

[1] Aaron Wright & Primavera De Filippi, Decentralized Blockchain Technology and The Rise of *Lex Cryptographia* 2 (Mar. 20, 2015, revised July 25, 2017) (unpublished manuscript), https://ssrn.com/abstract=2580664. *See also* John W. Bagby, David Reitter & Philip Chwistek, An Emerging Political Economy of the BlockChain: Enhancing Regulatory Opportunities (2018), https://ssrn.com/abstract=3299598.

[2] *See* Valentina Gatteschi, Fabrizio Lamberti, Claudio Demartini, Chiara Pranteda & Victor Santamaria, *To Blockchain or Not to Blockchain: That Is The Question,* IT Professional 62 (March/April 2018), IEEE Computer Society.

### Advantages

- Implements a shared repository that is maintained by peers—everyone can access data and view transactions. Moreover, storing information on nodes prevents data loss in case of unexpected events.
- Provides trust between parties. Digital signature and validation ensure that every node and user behaves correctly, without needing intermediaries.
- Could become a worldwide data repository accessed by different actors. Everyone can potentially read/write on it.
- Transparency is guaranteed. Everyone could read not only the final state of transactions, but also the history of passed states.
- Immutability. Data cannot be erased or changed.
- Decentralization. It can run without a central authority and cannot be controlled, censored, or shut down.
- Automation. With smart contracts, activities could be automatized.

### Disadvantages

- Characterized by high power consumption. A Bitcoin transaction could cost $6 when considering the energy consumed by network nodes.
- Mining requires expensive hardware, and the majority of computing power is wasted. Mining blocks is a competition among nodes where only the quickest wins—the others are just wasting resources. To increase the probability of winning, nodes could join mining pools and collaborate with other nodes, sharing revenues. A solution to reduce the amount of necessary computing power could be to change the mining process from proof of work to proof of stake, where nodes can purchase the opportunity to mine using tokens, and mining power is proportional to the number of tokens owned. This way, mining would be less resource intensive but would be restricted to token holders.
- Data replication requires space. Local copies of the blockchain (hence, of all transactions that have occurred since its creation—about 105 Gbytes for Bitcoin and 70 Gbytes for Bitcoin and Ethereum; http://bitinfocharts.com) are stored on each network node. Performances are therefore not yet comparable with databases.
- Adding information is slow. Creating a Bitcoin block takes around 10 to 60 minutes (http://blockchain.info/charts/avg-confirmation-time). Ethereum requires 15 seconds, (http://etherscan.io/chart/blocktime), a smaller though still significant amount of time.
- Immutability and transparency could harm users' privacy and reputation. Every network node would store a copy of the blockchain and could possibly access its content.
- Smart contracts cannot rely on external APIs. Every node should be able to process previous transactions and end with the same result as the other nodes. That is, information must be immutable. Consequently, data required by a smart contract should be first injected in the blockchain.

Oracles can enable this injection, but require a strong reputation system or governance mechanism and need to be as robust as the blockchain itself, not to become the weakest part of the process.

Smart contracts can be buggy. Because their code is publicly available and they become autonomous entities once they are created, they could be "candy for hackers." As they are stored on the blockchain, smart contracts cannot be modified. To remove code bugs, developers have to create new contracts and transfer all data and pointers from the old to the new ones. The most relevant case of a smart-contract-based attack happened on Ethereum in June 2016, when about $60 million was "stolen."[3]

### III. THE MECHANICS

#### A. The Basics

Blockchain is a modification and conglomeration of existing technology and concepts. Michael Scott explains, "The blockchain is a testament to the power of a single cryptographic primitive–the hash function. Really nothing else is required, so if you can get your head around the hash function, you can understand the basics of the blockchain."[4] Mr. Scott describes a hash by stating:

> A cryptographic hash function takes one input and calculates one output. For example, for the input 'We hold these truths to be self-evident', the well known hash function SHA256 produces the output:
>
> 84ba74b2661c87470665a1a5f5ab526afcf266f8c5effb795bef2d2514a8afd3
>
> For the slightly different input "we hold these truths to be self-evident" (note the lower case w), the output is
>
> 246160c031a4ddd9d940e931721fdec7e72087c8eccf5ea5621bb15d22959c19[5]

The above examples provide information about hash functions. Mr. Scott writes:

> The output bears no obvious relationship to the input, indeed it looks completely random. A tiny change to the input produces a completely different output . . . given just the output it's impossible to determine the input. For this reason the hash function is often called a "one way" hash function. Also, it's impossible to find two different inputs which give the same output. For the function SHA256, the 256 refers to the fact that the output is always the same length (actually 256 bits), independent of the length of the input.[6]

---

[3] *Id.* at 68.
[4] *See* Michael Scott, *The Essence of the Blockchain,* 1 (unpublished manuscript) (on file with authors).
[5] *Id.*

Blockchain gets its names from the chaining of the hash. Mr. Scott provides us with a diagram appearing here as Exhibit 1.[7]

**Exhibit 1**
**A Simple Hash Chain**

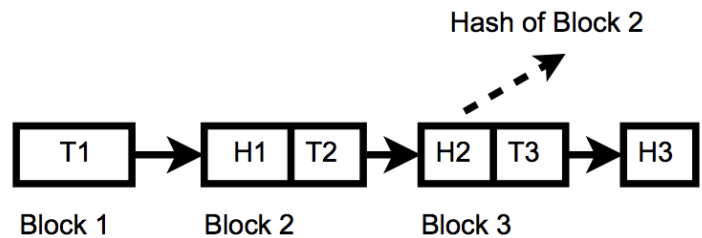

Mr. Scott writes the following explanation:

> Here the T are "transactions" of some sort. Examine this diagram for a while, and appreciate the power of the chaining. The value H3 is calculated by hashing the whole of block 3, which includes the hash of block 2, which in turn includes the hash of block 1 etc. Note that because of the one-wayness of the hash function, this chain can only be calculated from left-to-right. So already we have some of the properties we want. This hash chain can potentially be used as an immutable record of transactions. Any attempt to tamper with it can be detected, as the hashes will change.[8]

As Scott observes, "the term block comes a group of transactions. An important point–what we have described as individual transactions will probably consist in fact of a large batch of transactions all included inside of a single block.[9]

### B. Consensus Mechanism

The consensus mechanism is the crux of blockchain technology. Professor Okan Arabaci provides the following description:

---

[6] *Id.*
[7] *Id.* at 2.
[8] *Id.*
[9] *Id.*

> Consensus is the word for a general agreement or majority of opinion between a number of subjects. . . . Consensus in a blockchain network refers to the process where the distributed nodes agree on the history and the final state of the data on the ledger, usually referred to as distributed consensus. Since all participants in the network hold the data, they can also be a part of the decision-making. Every new block that gets added to the blockchain needs to be agreed upon according to the defined protocol so that the replication is done uniformly.[10]

There are numerous different consensus algorithms that have been developed. For example, Professors Deepak Puthal, Nisha Malik, Saraju P. Mohanty, Elias Kougianos, and Gautam Das describe the three major algorithms as follows:

### 1. Practical Byzantine Fault Tolerance Algorithm

> The practical Byzantine fault tolerance (PBFT) algorithm was proposed as a solution to the Byzantine Generals' Problem, which is about conducting a successful attack on a rival city by the Byzantine army. For the Byzantine army to win, all of the loyal generals must work from the same plan and attack simultaneously. In addition, no matter what the traitors do, the loyal generals should stick to the decided plan, as a small number of traitors could ruin the plan. Similarly, in the blockchain, PBFT works to establish consensus among the participating nodes. The nodes maintain their current state, and, when a new message is received, the current state and the message are fed together for computations to help the node reach a decision. This decision is then broadcast to the network. A majority of the decisions determines the consensus for the network. Hyperledger, which is working on developing consortium blockchain systems for businesses, utilizes PBFT as its underlying consensus mechanism. It should be pointed out that many of the new developments on blockchain stem from prior work on distributed databases.

### 2. Proof of Work

> Proof of work was the first decentralized consensus protocol proposed by Nakamoto to achieve consistency and security in the bitcoin network. In bitcoin, currency transfer occurs in a completely decentralized fashion, thus requiring a consensus for authentication and block validation. The nodes in the bitcoin network compete to calculate the hash value of the next block, which is supposed to be less than a dynamically varying target value, determined by the consensus rule. Nodes achieving the solution wait for confirmation by other nodes before adding the block to the existing blockchain. More than one valid block might be generated if multiple nodes find an appropriate solution causing a temporary fork (branch) in the network. In such scenarios, all of them are acceptable, and the nodes closer to the miners accept the solution they receive and forward the same to other peers. Conflict at a later stage is avoided by accepting the longest version of the chain available at any time.

---

[10] *See* Okan Arabaci, *Blockchain Consensus Mechanisms−The Case of Natural Disasters,* 17 Uppsala Universitet, ISSN: 1650-8319, UPTEC STS 18028 (July 2018).

### 3. Proof of Stake

Proof of stake was proposed to overcome the disadvantages of excessive power consumption by proof of work in bitcoin. Ethereum utilizes proof of stake to achieve consensus. Instead of investing in resources that can perform rigorous computations for hash calculations in proof of work, proof of stake proposes to buy cryptocurrency and use it as stake in the network. The stake is directly proportional to the chance of becoming the block validator. To reach consensus, the block validator is randomly selected and is not predetermined. The nodes producing valid blocks get incentives, but, if their block is not included in the existing chain, they also lose some amount of their stake.[11]

Professors Deepak Puthal, Nisha Malik, Saraju P. Mohanty, Elias Kougianos, and Gautam Das, discuss the following factors to be considered as various consensus models are being developed:

• Type of blockchain: A blockchain network can be permissioned or permissionless.
• Transaction rate: The consensus algorithm basically decides the rate at which transactions are confirmed. In bitcoin, which employs proof of work, the transaction rate is only seven transactions/s, because proof of work requires significant computation time and the block generation time is 10 min.
• Scalability: A blockchain system is scalable if it can achieve consensus with the number of nodes continuously growing, especially in public blockchain systems.
• Participation charges: For some systems, an initial cost of participation is required. For example, with proof of stake, nodes invest in the cryptocurrency to express their interest in the consensus and block validation, whereas proof of work requires energy input, which is not necessary if you simply want to be part of the network and do not wish to mine.
• Trust condition: This determines whether the nodes contributing are to be trusted and predetermined (as in consortium and private blockchain

---

[11] *See* Deepak Puthal, Nisha Malik, Saraju P. Mohanty, Elias Kougianos & Gautam Das, *Everything You Wanted to Know About the Blockchain: Its Promise, Components, Processes, and Problems,* 6 (July 2018) (unpublished Manuscript) (on file with authors).

systems) or unknown (as in public and proof of work-based blockchains).[12]

### C. Types of Blockchains

Blockchains can either be permissionless (public) or permissioned (private).[13] Karl Wüst describes the differences:

#### Permissionless Blockchains

Bitcoin and Ethereum are instances of permissionless blockchains, which are open and decentralized. Any peer can join and leave the network as reader and writer at any time. Interestingly, there is no central entity which manages the membership, or which could ban illegitimate readers or writers. This openness implies that the written content is readable by any peer. With the use of cryptographic primitives however, it is technically feasible to design a permissionless blockchain which hides privacy relevant information (e.g. Zerocash).

#### Permissioned Blockchains

To only authorize a limited set of readers and writers, so called-permissioned blockchains have been recently proposed. Here, a central entity decides and attributes the right to individual peers to participate in the write or read operations of the blockchain. To provide encapsulation and privacy, reader and writer could also run in separated parallel blockchains that are interconnected. The most widely known instance of permissioned blockchains are Hyperledger Fabric and R3 Corda.[14]

### D. Smart Contracts

An integral part to the disruptive influence of blockchain is smart contracts. Bhabendu Kumar Mohanta, Soumyashree S. Panda, and Debasish Jena, describes smart contracts as:

a computer program having self verifying, self-executing, tamper-resistant properties. The smart contract concept was proposed by Nick Szabo in 1994. It allows executing code without the third parties. A smart contract consists of the value, address, functions, and state. It takes transaction as an input, executes the corresponding code and triggers the output events. Depending upon the function logic implementation states are changes. . . . Some characterizes of a smart contract are:

---

[12] *Id.*
[13] *See* Karl Wüst & Arthur Gervais, *Do You Need a Blockchain?,* 2018 Crypto Valley Conference on Blockchain Technology, 45 IEEE Computer Society, DOI 10.1109/CVCBT.2018.00011.
[14] *Id.* at 46.

- Smart contract are machine readable code run on blockchain platform
- Smart contracts are part of one application program
- Smart contracts are event driven program
- Smart contracts are autonomous once created no need to monitor
- Smart contracts are distributed[15]

These abilities to perform agreements at machine speed and emulate an escrow without a need for a trusted third party, provide vast capabilities as seen in the use cases in Section VIII (*Infra.*).

## IV. ANALYSIS OF TWO DISTRIBUTED LEDGER TECHNOLOGIES[16]

### A. IOTA

Named after the ninth letter of the Greek alphabet, the IOTA protocol is an open-source distributed ledger technology developed by the IOTA Foundation, and designed denovo to power the Machine Economy through data integrity and fee-less micro-transactions.[17] Bitcoin (BTC) and IOTA are both distributed ledger technologies. Bitcoin technology utilizes a vanilla version of blockchain technology where transactions are combined together into a block along with the hash of the previous block.[18] This creates an irreversible and unchangeable chain of groups (blocks) of transactions.[19] IOTA utilizes a directed acyclic graph (DAG or Tangle) instead of this linear chaining[20] (both can be seen in Exhibit 2). In IOTA, nodes or members comprise a self-rewarding network which verifies transactions in a non-linear manner.[21] IOTA also has irreversible and unchangeable transactions along with Bitcoin's decentralized system. IOTA, however, does not require mining (the computationally heavy requirement for transaction verification) nor does it employ transaction fees.[22]

---

[15] *See* Bhabendu Kumar Mohanta, Soumyashree S. Panda & Debasish Jena, *An Overview of Smart Contract and Use Cases in Blockchain Technology,* IEEE – 43488 (2018).
[16] Bitcoin is described in great detain at Section V. (*Infra*).
[17] *See* IOTA, What is IOTA?, https://www.iota.org/get-started/what-is-iota.
[18] *See* Michael Scott, *supra* note 4 at 3.
[19] *See* Michael Scott, *supra* note 4 at 4.
[20] *See* Nihat Yuva & İsmail Kırbaş, *Directed Acyclic Graph Based on Crypto Currency Application Example: IOTA,* Proceedings of 1st International Conference on Data Science and Applications (ICONDATA'18), Yalova, TURKEY on Oct. 4-7, 2018. (on file with authors).
[21] *See* Michael J. Casey & Paul Vigna, The Truth Machine: The Blockchain and the Future of Everything. An Introduction to Cryptography and Cryptocurrencies 131-32 (St. Martin's Press, 2018).
[22] *See* Nihat Yuva & İsmail Kırbaş, *Directed Acyclic Graph Based on Crypto Currency Application Example: IOTA,* Proceedings of 1st International Conference on Data Science and Applications

**Exhibit 2**
**Underlying Data Structure of IOTA and Bitcoin**[23]

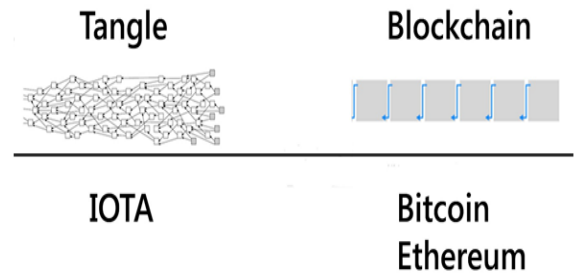

### B. Advantages and Disadvantages

BTC advantages include its wide spread adoption and testing, direct payments with no intermediaries, low fees compared to traditional transaction methods, and support for basic smart contracts (Such as OP_CHECKLOCKTIMEVERIFY which makes a transaction not spendable until verified). The disadvantages of BTC include that the consensus mechanism relies on a majority vote and thus is vulnerable to attack if one entity controls the majority of the system. In addition: the wallet (a person's identity in the system) can be lost; there is no built-in mechanism for refunds; and the consensus mechanism is proof of work based, creating high waste of electricity and computing power.[24]

IOTA advantages include: no transaction fees; more scalable than Bitcoin; quantum proof by design; and IOTA requires no mining (transaction verification). The disadvantages of IOTA at present include: scalability requiring lots of users; and unnecessary and vulnerable hash function; and the lack of smart contracts (still underdeveloped).[25]

---

(ICONDATA'18), Yalova, TURKEY on Oct. 4-7, 2018. (on file with authors).
[23] *See* Hackermoon (graphic); Nihat Yuva & İsmail Kırbaş, *Directed Acyclic Graph Based on Crypto Currency Application Example: IOTA,* Proceedings of 1st International Conference on Data Science and Applications (ICONDATA'18), Yalova, TURKEY on Oct. 4-7, 2018. (on file with authors) (technical citation behind the graphic).
[24] *See* George Cornel Dumitrescu, *Bitcoin− A Brief Analysis of the Advantages and Disadvantages,* http://www.globeco.ro/wp-content/uploads/vol/split/vol_5_no_2/geo_2017_vol5_no2_art_008.pdf.
[25] *See* Michael Colavita & Garrett Tanzer, *A Cryptanalysis of IOTA's Curl Hash Function,* (unpublished manuscript) (on file with authors).

### C. Overall Analysis

In a technical sense, Bitcoin is a safer option over IOTA since it relies on very basic principles and well-proven/tested algorithms.[26] IOTA employs the use of a custom hash function, not standard use in the industry, and thus introduces additional security risk from a technical perspective.[27] Bitcoin, on the other hand, employs the industry standard hash function.[28] Additionally, Bitcoin's protocols are well tested due to the publicity and the drastic value volatility the coin has experienced recently.[29] IOTA protocols rely on a more complex concept (of directed acyclic graphs) which is less tested than Bitcoin's; however, IOTA's protocols claim to be quantum computing resistant.[30]

In September of 2017, a group of researchers published a paper identifying potential vulnerabilities in the underlying hash function to IOTA.[31] IOTA has since released statements that claiming that the paper disregarded several key factors of the IOTA protocol which mitigates the identified vulnerabilities.[32] IOTA ended up swapping out their custom hash function for an industry standard hash function.[33] The strength of IOTA is still not clear,[34] but if IOTA's claim to be quantum computing resistant proves accurate, this additional security strength may prove to be a highly valuable attribute.

---

[26] *See* Arvind Narayanan, Joseph Bonneau, Edward Felten, Andrew Miller & Steven Goldfed, *Bitcoin and Cryptocurrency Technologies: A Comprehensive Introduction,* Ch. 1, An Introduction to Cryptography and Cryptocurrencies, Princeton U. Press, 2016).

[27] *See* Colavita & Tanzer, *supra* note 24.

[28] *See* Arvind Narayanan, Joseph Bonneau, Edward Felten, Andrew Miller & Steven Goldfed, *Bitcoin and Cryptocurrency Technologies: A Comprehensive Introduction,* Ch. 1, An Introduction to Cryptography and Cryptocurrencies, Princeton U. Press, 2016).

[29] See Lawrence J. Trautman, Bitcoin, Virtual Currencies and the Struggle of Law and Regulation to Keep Pace, 102 MARQ. L. REV. 447 (2018), https://ssrn.com/abstract=3182867.

[30] *See* Nihat Yuva & İsmail Kırbaş, *Directed Acyclic Graph Based on Crypto Currency Application Example: IOTA,* Proceedings of 1st International Conference on Data Science and Applications (ICONDATA'18), Yalova, TURKEY on Oct. 4-7, 2018. (on file with authors).

[31] *See* Ethan Heilman, Neha Narula, Thaddeus Dryja & Madars Virza, *IOTA Vulnerability report: Cryptanalysis of the Curl Hash Function Enabling Practical Signature Forgery Attacks on the IOTA Cryptocurrency,* GITHUB (Sept. 7, 2017), https://github.com/mit-dci/tangled-curl/blob/master/vuln-iota.md.

[32] *See Official IOTA Foundation Response to the Digital Currency Initiative at the MIT Media Lab− Part 4 / 4,* IOTA Foundation, (Jan. 7, 2018), https://blog.iota.org/official-iota-foundation-response-to-the-digital-currency-initiative-at-the-mit-media-lab-part-4-11fdccc9eb6d.

[33] *See* Morgan Peck, *Cryptographers Urge People to Abandon IOTA After Leaked Emails,* IEEE Spectrum (Feb. 27, 2018), https://spectrum.ieee.org/tech-talk/computing/networks/cryptographers-urge-users-and-researchers-to-abandon-iota-after-leaked-emails.

[34] *See* Heilman, et al., *supra* note 30.

### D. Impact of Quantum Computing

A robust body of scholarly literature has been written about quantum computing.[35] For example, a report sponsored by the US Department of Energy provides our working definition, "Quantum computing uses computational elements that obey quantum mechanical laws to potentially provide transformative changes in computational power for certain problems of interest."[36] One such instance of a breakthrough is "Peter Shor's 1994 breakthrough discovery of a polynomial time quantum algorithm for integer factorization [which] sparked great

---

[35] *See* Nicolae Bulz, Alexandru Bogdan, Sorin Chelmu & Amalia Strateanu, Biodiversity 'And' Systemic Knowledge Engineering – Theory, Logic and Praxis (I) – An Interdisciplinary Way Toward Knowledge Society/Integrated Project for the Exergy and Sustainable Development of the Agro-Biodiversity Through an Interactive Modeling - Analysis Regarding the Synergy between Rural Ecoeconomy and Bioethics Based Upon Smart Growth, Eco-Innovation and Large Scale Systems (2011), In RECENT RESEARCH IN ENERGY, ENVIRONMENT, DEVICES, SYSTEMS, COMMUNICATIONS, COMPUTERS, WSEAS Press, 96 https://ssrn.com/abstract=2383685; Joanna Caytas, *Directionality of Time in Quantum Computing*, 10(1) COLUM. SCI. REV. 8 (Fall 2013), https://ssrn.com/abstract=2368826; Marcos López de Prado, Quantum Computing (in 5 minutes or Less, Berkeley Lab, Lawrence Berkeley National Laboratory, (presentation Slides) (2015), https://ssrn.com/abstract=2694133; Carlos Pedro dos Santos Gonçalves, Quantum Robotics, Neural Networks and the Quantum Force Interpretation (2018), https://ssrn.com/abstract=3244327; David Ellerman, *Quantum Mechanics Over Sets,* (2013), https://ssrn.com/abstract=2344945; Jon Lindsay, Why Quantum Computing Will Not Destabilize International Security: The Political Logic of Cryptology (2018), https://ssrn.com/abstract=3205507; Stefan Heng, *Rising Stars in Information and Communication Technology,* Deutsche Bank Research Economics Working Paper No. 46. (2004), https://ssrn.com/abstract=620281; Yogesh Malhotra, Advancing Cognitive Analytics Using Quantum Computing for Next Generation Encryption (Presentation Slides) (2017), https://ssrn.com/abstract=2940467; Michael Marzec, Portfolio Optimization: Applications in Quantum Computing (2014), https://ssrn.com/abstract=2278729; Gili Rosenberg, Poya Haghnegahdar, Phil Goddard, Peter P. Carr, Kesheng Wu & Marcos López de Prado, *Solving the Optimal Trading Trajectory Problem Using a Quantum Annealer,* IEEE JOURNAL OF SELECTED TOPICS IN SIGNAL PROCESSING, (2016), https://ssrn.com/abstract=2649376; Neeraj Samtani, *How Would Quantum Computing Impact the Security of Bitcoin by Enhancing Our Ability to Solve the Elliptic Curve Discrete Logarithm Problem?,* (2018), https://ssrn.com/abstract=3232101; P. H. Sureshkumar, Ambily Pramitha & Rajesh Ramachandran, *The Quantum Key Distribution (QKD) Based Security Enhanced Cloud Data Center Connectivity*, 7(4) INT'L J. LATEST TRENDS IN ENGINEERING AND TECH. (2018), https://ssrn.com/abstract=3128681; Florenta Teodoridis, Startup Commercialization Strategies of Disruptive Technologies: Implications for the Rate of Scientific Discovery (2017), https://ssrn.com/abstract=3062776; Andre van Tonder, A Lambda Calculus for Quantum Computation, 2003(7) Comp. Sci. Preprint Archive 75 (2003), https://ssrn.com/abstract=2978398; Craig S. Wright, Bitcoin and Quantum Computing (2017), https://ssrn.com/abstract=3152419; Vyacheslav I. Yukalov & Didier Sornette, Scheme of Thinking Quantum Systems, 6 Laser Physics Letters 833 (2009), https://ssrn.com/abstract=1470624; Wen-Ran Zhang & Karl Ernest Peace, *Causality Is Logically Definable — Toward an Equilibrium-Based Computing Paradigm of Quantum Agents and Quantum Intelligence (QAQI) (Survey and Research),* 4 J. QUANTUM INFO. SCI. 227 (2014), https://ssrn.com/abstract=2541845.

[36] *See* Alán Aspuru-Guzik, Wim van Dam, Edward Farhi, Frank Gaitan, Travis Humble, Stephen Jordan, Andrew Landhal, Peter Love, Robert Lucas, John Preskill, Richard Muller, Krysta Svore, Nathan Wiebe, Carl Williams, Ceren Susut, ASCR REPORT ON QUANTUM COMPUTING FOR SCIENCE 2 (2015).

interest in discovering additional quantum algorithms and developing hardware on which to run them."[37]

RSA is perhaps the most widely used public key cryptography algorithms.[38] Professors Hoffstein, Pipher and Silverman highlight the importance of integer factorization for RSA, observing "the security of RSA relies on the apparent difficulty of factoring large numbers."[39] The DOE reports:

> quantum computing offers a fundamentally new approach to computation that promises capabilities not available with today's existing transistor-based processing. So far, the theory of quantum computing has found significant speed-ups to a few prominent algorithms in modeling, simulation and mathematics, and experimental efforts in quantum computer science have recently made great strides demonstrating crude quantum algorithms to solve modest problems in physical simulation and applied mathematics. In addition, it is believed that the operation of an idealized 100-qubit quantum computer may exceed the simulation capabilities of even future exascale computers. This suggests that quantum computers may have the potential to enable some aspects of computational science to progress far beyond exascale.[40]

A report by the MITRE corporation, a Federally Funded Research and Develop Center (FFRDC) for the US Government, examines the potential impact of quantum computing to blockchain technology:

> The computational data structure known as a blockchain provides an open, public, distributed ledger that has many interesting applications, including digital currencies. The security of this ledger depends on the difficulty of solving certain cryptographic problems which are undermined by the potential of quantum computation. Specifically, hashes as used in signing the blocks of the ledger can be compromised, as can any public/private key system which relies on the so called hidden subgroup problem.[41]

Brandon Rodenburg and Stephen P. Pappas continue to provide the following analysis:

> In the context of quantum computing, we are confronted with two aspects of invalidating the promises of blockchain. First, the inversion of hashes is assumed to be computationally difficult. If this can be dramatically simplified by a quantum computer, the authenticity of the upstream blockchain can no longer be guaranteed and the authenticity of entries in the blockchain is compromised. . . . As a secondary threat, in any aspect of a blockchain implementation that uses public/private key cryptography, whether it be in information exchange between parties or in digital signatures, a quantum computer may be able to break the security of the encryption.[42]

---

[37] *Id.*
[38] *See* Shireen Nisha & Mohammad Farik, *RSA Public Key Cryptography Algorithm−A Review,* 6(7) INT'L J. SCI & TECH. RES. 187 (2017).
[39] *See* JEFFREY HOFFSTEIN, JILL PIPHER & JOSEPH H. SILVERMAN, AN INTRODUCTION TO MATHEMATICAL CRYPTOGRAPHY 133 (Springer, 2008).
[40] *See* Aspuru-Guzik, et al., *supra* note 35 at 6.
[41] *See* Brandon Rodenburg & Stephen P. Pappas, *Blockchain and Quantum Computing,* MITRE Corporation MTR170487 Mitre Tech. Rpt. v (2017).

If IOTA's claim to be quantum computing resistant proves accurate, then, even as quantum computing capabilities are developed, there should be no additional security risk.[43] This is not the case for Bitcoin's protocols since they rely on industry standard algorithms which are not quantum resistant and thus would require quick adaptation and presents a security risk as quantum computing capabilities continue to become operational. IOTA's protocols also allow for dramatically more transactions to occur and be verified in a shorter period of time.[44]

From a business perspective, both technologies offer substantial abilities. Bitcoin's strength is its wide-spread use and adoption. This adoption allows for easier integration into existing systems and uses of Bitcoin technology. Bitcoin being one of the first distributed ledger technologies allows for easier software development and integration into systems. However, IOTA's feeless and scalable design allows for a more realistic replacement of existing systems. A summary comparison can be seen in the graphic in Exhibit 4.

### E. Challenges

The biggest problem that distributed ledger technologies face, are scalability for adoption into the real world.[45] Even cryptosystems like Bitcoin, which have risen to public notoriety, lack the scalability to be adopted in a way which largely replaces current systems.[46] For example, as Bitcoin now exists, it cannot be adopted by a country or large bank to replace the existing online banking systems because of its limitations as to: (1) the number of transactions which can be processed at a time; and (2) the prolonged time it takes to verify a transaction.[47] A number of modifications have been developed to speed up the transaction and verification rate for Bitcoin, but these modifications are not enough for wide spread use or adoption.[48] An extensive overhaul of the Bitcoin protocols would be necessary in order for it to function on a wide-spread system. This lack of scalability in the design of Bitcoin creates a challenge for realistic use while IOTA's inherent design for scalability makes IOTA a far more attractive choice.[49] A depiction of scalability of Bitcoin and IOTA can be seen in Exhibit 3.

---

[42] *Id.* at 4.
[43] *See* Nihat Yuva & İsmail Kırbaş, *Directed Acyclic Graph Based on Crypto Currency Application Example: IOTA,* Proceedings of 1st International Conference on Data Science and Applications (ICONDATA'18), Yalova, TURKEY on Oct. 4-7, 2018. (on file with authors).
[44] *Id.*
[45] Kyle Croman, Christian Decker, Ittay Eyal, Adem Efe Gencer, Ari Juels, Ahmed Kosba, Andrew Miller, Prateek Saxena, Elaine Shi, Emin Gün Sirer, Dawn Song & Roger Wattenhofer, On Scaling Decentralized Blockchains (unpublished ms.) (on file with authors).
[46] *Id.*
[47] *Id.*

**Exhibit 3**
**Scalability of Bitcoin & IOTA**[50]

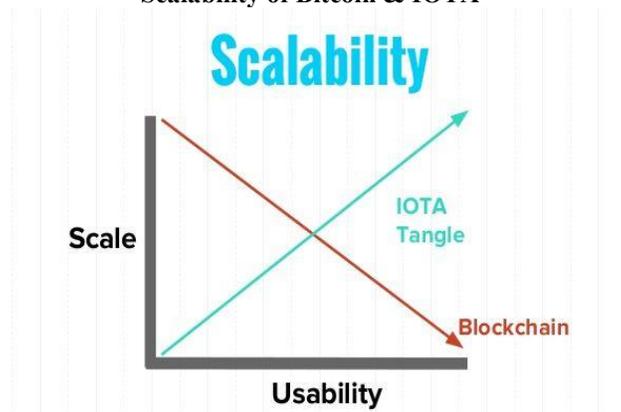

IOTA's biggest challenge is development and adoption.[51] IOTA's strength is in its scalability and functions optimally with numerous members. Yet, the full use of IOTA's technology is still being developed before this wide adoption can occur.[52] A big hurtle being tackled for adoption is the development of smart contracts.[53] IOTA naively supports smart contracts but the actual overlying software which makes use of these abilities is not yet developed as some frameworks are for Bitcoin.[54] A summary of the transactional difference between IOTA and BTC are detailed in Exhibit 4, below.

---

[48] *Id.*
[49] *See* Nihat Yuva & İsmail Kırbaş, *Directed Acyclic Graph Based on Crypto Currency Application Example: IOTA,* Proceedings of 1st International Conference on Data Science and Applications (ICONDATA'18), Yalova, TURKEY on Oct. 4-7, 2018. (on file with authors).
[50] *Id.*
[51] *Id.*
[52] *See* David Sønstebø, IOTA Year in Review, Year in Preview, IOTA, https://blog.iota.org/year-in-review-year-in-preview-85686fe1bb3b.
[53] *See* Nihat Yuva & İsmail Kırbaş, *Directed Acyclic Graph Based on Crypto Currency Application Example: IOTA,* Proceedings of 1st International Conference on Data Science and Applications (ICONDATA'18), Yalova, TURKEY on Oct. 4-7, 2018. (on file with authors).
[54] *Id.*

**Exhibit 4**
**Comparison of IOTA and BTC**[55]

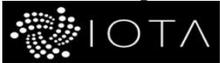

### F. When to Use Blockchain

The previously mentioned capabilities, benefits, and costs calls into question when blockchain should be used. Concrete examples are described below in Section VIII (*Infra*). Karl Wüst and Arthur Gervais provide the following insight into when blockchain may be reasonably applied "In general, using an open or permissioned blockchain only makes sense when multiple mutually mistrusting entities want to interact and change the state of a system, and are not willing to agree on an online trusted third party. . . ."

If no data needs to be stored, no database is required at all, i.e. a blockchain, as a form of database, is of no use. Similarly, if only one writer exists, a blockchain does not provide additional guarantees and a regular database is better suited, because it provides better performance in terms of throughput and latency. If a trusted third party (TTP) is available, there are two options. First, if the TTP is always online, write operations can be delegated to it and it can function as verifier for state transitions. Second, if the TTP is usually offline, it can function as a certificate authority in the setting of a permissioned blockchain, i.e. where all writers of the system are known. If the writers all mutually trust each other, i.e. they assume that no participant is malicious, a database with shared write access is likely the best solution. If they do not trust each other, using a permissioned blockchain makes sense. Depending on whether public verifiability is required, anyone can be allowed to read the state (public permissioned blockchain) or the set of readers may also be restricted (private permissioned blockchain). If the set of writers is not fixed and known to the participants, as is the case for many

---

[55] *Id.*

cryptocurrencies such as Bitcoin, a permissionless blockchain is a suitable solution. . . .

In a centralized system, the performance in terms of latency and throughput is generally much better than in blockchain systems, as blockchains add additional complexity through their consensus mechanism. For example, Bitcoin can currently only sustain a throughput of approximately seven transactions per second (which could be extended to approximately 66 without compromising security), while a centralized system such as Visa can handle peaks of more than fifty thousand transactions. There is a tradeoff between decentralization, i.e. how well a system scales to a large number of writers without mutual trust, and throughput, i.e. how many state updates a system can handle in a given amount of time. When making the decision of whether to use a blockchain system or not, this tradeoff should be taken into account as well.[56]

These tradeoffs between permissionless and permissioned blockchains are summarized in Exhibit 5 below:

**Exhibit 5**
**Comparison of Permissionless and Permissioned Blockchains[57]**

|  | Permissionless Blockchain | Permissioned Blockchain | Central Database |
|---|---|---|---|
| Throughput | Low | High | Very High |
| Latency | Slow | Medium | Fast |
| Number of readers | High | High | High |
| Number of writers | High | Low | High |
| Number of untrusted writers | High | Low | 0 |
| Consensus mechanism | Mainly PoW, some PoS | BFT protocols (e.g. PBFT [5]) | None |
| Centrally managed | No | Yes | Yes |

Professors Karl Wüst and Arthur Gervais provide a decision tree that considers above statements seen in Exhibit 6:

---

[56] *See* Wüst & Gervais, *supra* note 13 at 46.
[57] *Id.* at 48.

**Exhibit 6**
**Blockchain Decision Tree**[58]

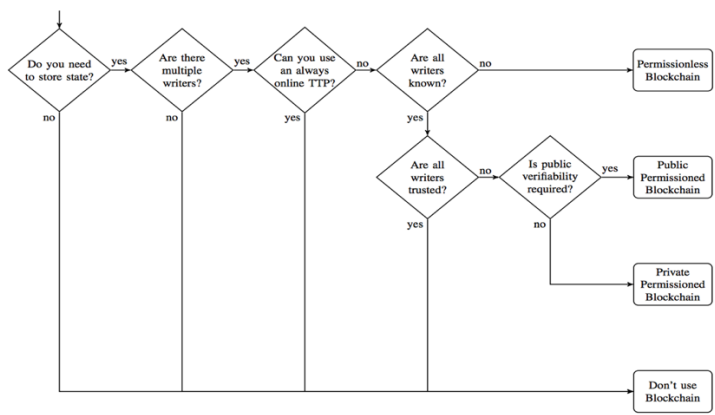

## V.  VIRTUAL CURRENCIES

Despite the volatility and precipitous decline in the value of virtual currencies during 2018, it is the hype and price performance of this asset class and particularly Bitcoin that accounts for the rapid global awareness of the blockchain since 2009. *The Wall Street Journal* observes, "After rising nearly 1,400% in 2017, bitcoin reversed hard in 2018, falling about 70% and erasing some $160 billion worth of value."[59] Financial journalist Paul Vigna asks, "Where does the digital currency go now that many speculators have been wiped out? . . . The selloff cost the crypto market $700 billion in 2018, dwarfing the $15.7 billion raised by initial coin offerings and the $2.6 billion venture-capital firms invested in crypto startups."[60]

The U.S. Department of the Treasury, FinCEN division, has defined a virtual currency as, "those currencies that operate like a currency in some environments, but does not have legal tender status in any jurisdiction."[61] The Financial Action Task Force provides a more comprehensive virtual currency definition:

---

[58] *Id.* at 47.
[59] *See* Paul Vigna, *Crypto Investors Dig Out of Disaster,* WALL ST. J., Jan. 2, 2019 at B10.
[60] *Id.*
[61] Lawrence Trautman, *Virtual Currencies; Bitcoin & What Now After Liberty Reserve, Silk Road, and Mt. Gox?*, 20 RICH. J. L. & TECH. 1, 3 (2014) [hereinafter Trautman, *Virtual Currencies*] (citing *Beyond Silk Road: Potential Risks, Threats, and Promises of Virtual Currencies: Hearings Before the S. Comm. on Homeland Security & Governmental Affairs*, 113th Cong. 5–6 (2013) (statement of Jennifer Shasky Calvery, Director, Fin. Crimes Enforcement Network, U.S. Dep't of the Treasury)); *accord* DEP'T OF TREASURY, FIN. CRIMES ENF'T NETWORK, FIN-2013-G001, APPLICATION OF FINCEN'S REGULATIONS TO PERSONS ADMINISTERING, EXCHANGING, OR USING VIRTUAL CURRENCIES (2013) [hereinafter FIN-2013-G001].

a digital representation of value that can be digitally traded and functions as (1) a medium of exchange; and/or (2) a unit of account; and/or (3) a store of value, but does not have legal tender status (i.e., when tendered to a creditor, is a valid and legal offer of payment) in any jurisdiction. It is not issued or guaranteed by any jurisdiction, and fulfils the above functions only by agreement within the community of users of the virtual currency. Virtual currency is distinguished from fiat currency (a.k.a. "real currency," "real money," or "national currency"), which is the coin and paper money of a country that is designated as its legal tender; circulates; and is customarily used and accepted as a medium of exchange in the issuing country. It is distinct from e-money, which is a digital representation of fiat currency used to electronically transfer value denominated in fiat currency.[62]

### A. History

Trautman and Harrell have previously documented the history of barter, money, evolution of primitive money, development of a schematic for regulation of money in the United States, and modern approach to regulation and payment system mechanics.[63] We will not repeat that description here. Many observers trace the genesis of virtual currencies to David Chaum's 1982 crypto journal article.[64] Virtual assets and marketplaces found in Massively Multiplayer Online Games (MMOGs) such as *Second Life* where virtual assets were exchanged for actual sovereign currencies were also an inspiration for the development of cyber currencies.[65]

### B. Bitcoin

Then, in 2009, the creation of Bitcoin is credited to a pseudonymous hacker or group of hackers known as Satoshi Nakamoto.[66] For the first few years of dramatic growth in virtual currencies, the history of Bitcoin is materially synonymous with the widespread awareness of the blockchain.

---

[62] *See* FIN. ACTION TASK FORCE, VIRTUAL CURRENCIES: KEY DEFINITIONS AND POTENTIAL AML/CFT RISKS 4 (2014), https://www.fatf-gafi.org/media/fatf/documents/reports/Virtual-currency-key-definitions-and-potential-aml-cft-risks.pdf.
[63] *See* Lawrence J. Trautman & Alvin C. Harrell, *Bitcoin Versus Regulated Payment Systems: What Gives?*, 38 CARDOZO L. REV. 1041, 1053 (2017).
[64] *See* David Chaum, *Blind Signatures for Untraceable Payments*, *in* ADVANCES IN CRYPTOLOGY: PROCEEDINGS OF CRYPTO 82, 199 (David Chaum et al. eds., 1982).
[65] *See* Trautman & Harrell, *supra* note 63 at 1043.
[66] Lawrence J. Trautman, *Is Disruptive Blockchain Technology the Future of Financial Services?*, CONSUMER FIN. L.Q. REP. 232, 234 (2016).

### C. Cryptocurrency Universe

As of January 3, 2019, Coinmarketcap.com lists 2,083 different cryptocurrencies, having a total market capitalization of approximately $130.95 billion.[67] This contrasts with just 656 different cybercurrencies reported by Trautman and Harrell as of July 15, 2016, having a market capitalization approximating $13.01 billion.[68] Ranked by market capitalization at January 3, 2019, the top ten cybercurrencies are: Bitcoin ($67.26 billion); Ethereum ($15.6 billion); Ripple ($14.7 billion); Bitcoin Cash ($2.8 billion); EOS ($2.4 billion); Stellar ($2.1 billion); Litecoin ($1.9 billion); Bitcoin SV ($1.5 billion); Tron ($1.3 billion); and Cardano ($1.1 billion).[69] By far the largest of these more than two thousand cryptocurrencies, Bitcoin's market capitalization is comparable to 51.3% of all virtual currencies combined as of January 3, 2019.[70]

### D. International Acceptance Differs

The worldwide market for virtual currencies, and Bitcoin in particular, has encountered uneven acceptance during the past few years[71]—with industrialized countries such as Japan[72] and Germany[73] welcoming Bitcoin. However, regulators in other jurisdictions, such as South Korea,[74] Canada,[75] and China[76] have been much open to embracing virtual currencies to date.

---

[67] *All Cryptocurrencies*, COINMARKETCAP, HTTPS://COINMARKETCAP.COM/COINS/VIEWS/ALL/ (last visited Jan. 3, 2019). *See generally* Thomas Grove, *Farmer Banks on Cryptocurrency*, WALL ST. J., Apr. 23, 2018 at A9 (stating "Cryptocurrencies have sparked new interest in the possibilities of working outside the Russian financial system and the volatile ruble, which lost half of its value in 2014 amid a drop in oil prices"); Wilko Bolt & Maarten R.C. van Oordt, *On the Value of Virtual Currencies*, (Bank of Canada Working Paper No. 2016-42, 2016), https://ssrn.com/abstract=2767609.
[68] *See* Trautman & Harrell, *supra* note 62 at 1053.
[69] *All Cryptocurrencies*, *supra* note 66.
[70] *Id.*
[71] *See* Lawrence J. Trautman, *Bitcoin, Virtual Currencies and the Struggle of Law and Regulation to Keep Pace,* 102 MARQ. L. REV. 447, 475 (2018), https://ssrn.com/abstract=3182867, *citing* Matina Stevis-Gridneff & Georgi Kantchev, *Developing Countries Embrace Bitcoin*, WALL ST. J., Jan. 5, 2018, at B8.
[72] *See* Paul Vigna & Gregor Stuart Hunter, *Bitcoin Wins in Japan, Loses in South Korea*, WALL ST. J., Sept. 30–Oct. 1, 2017, at B9.
[73] *See* Zeke Turner, *Cash-Loving Germany Embraces Bitcoin*, WALL ST. J., Jan. 9, 2018 at B10.

### E. Criminal Use

During 2018, James Andrew Lewis, Senior Vice President at the Center for Strategic and International Studies (CSIS), provided Congressional testimony that "cybercrime is big business . . . [and] the last few years have shown that the Internet has a dark underside that is deeply troubling. The Internet has brought tremendous economic benefit, but this comes with the costs created by espionage and crime."[77] Mr. Lewis warns, "[t]he development of cryptocurrencies reduced risk and increased returns [to cybercriminals], by increasing the anonymity and ease of criminal transactions. The cybercrime monetization process is increasingly digitized, with criminals moving stolen funds rapidly among accounts with the goal of using it to buy cryptocurrencies in untraceable ways."[78]

### F. Anti-Money Laundering (AML) Considerations and FinCEN

The FinCEN is a small bureau of the U.S. Treasury Department, having only approximately 340 employees during 2013, and reports directly to the Office of Terrorism and Financial Intelligence. FinCEN has stated its' mission as:

> to safeguard the financial system from illicit use, combat money laundering and promote national security through the collection, analysis, and dissemination of financial intelligence and the strategic use of financial authorities."[79] Among FinCEN's responsibilities is to issue regulations and administer the Bank Secrecy Act (BSA).[80] The BSA requires that a wide range of financial institutions assist FinCEN by

---

[74] *See* Vigna & Hunter, *supra* note 71; Steven Russolillo & Kwanwoo Jun, *Seoul Acts Tough on Bitcoin*, WALL ST. J., Dec. 29, 2017, at B11; Eun-Young Jeong, *Korea Gets Tougher on Digital Currencies*, WALL ST. J., Jan. 9, 2018, at B10.
[75] *See* Paul Vieira, *Bitcoin Alarms Canadian Official*, WALL ST. J., Dec. 15, 2017, at B14.
[76] *See* Chao Deng & Paul Vigna, *China Cracks Down on Bitcoin*, WALL ST. J., Sept. 12, 2017, at A1; Chao Deng, *Beijing Draws a Bead on Virtual Currency*, WALL ST. J., Sept. 19, 2017, at B1; Steven Russolillo & Eun-Young Jeong, *Bitcoin Lures Asia Investors*, WALL ST. J., Dec. 13, 2017, at B1.
[77] *See After the Breach: The Monetization and Illicit Use of Stolen Data: Hearing Before the Subcomm. on Terrorism and Illicit Finance of the H. Comm. on Fin. Servs.*, 115th Cong. 2 (2018) (statement of James Andrew Lewis, Senior Vice President, Center for Strategic and International Studies (CSIS)) [hereinafter *After the Breach*]; *accord* Trautman, *Virtual Currencies*, *supra* note 60, at 8.
[78] *After the Breach*, *supra* note 76 at 3.
[79] *See* Trautman, *Virtual Currencies*, *supra* note 60 at 31 (citing *Beyond Silk Road: Potential Risks, Threats, and Promises of Virtual Currencies: Hearings Before the S. Comm. on Homeland Security & Governmental Affairs*, *supra* note 11, at 5 (statement of Jennifer Shasky Calvery, Director, Fin. Crimes Enforcement Network, U.S. Dep't. of the Treasury)).
[80] *Id.*

> having effective anti-money-laundering (AML) programs and by filing periodic reports with FinCEN and by maintaining appropriate records. Examples of these financial institutions include securities and futures broker/dealers, insurance companies, banks, casinos, other money services businesses, and certain trades or businesses such as automobile dealers.[81]

A comprehensive discussion of virtual currencies far exceeds the scope of this single article. However, many scholarly publications address this topic in great detail.[82]

For our purposes here, we will simply observe that during 2013, FinCEN Director, Jennifer Shasky Calvery states:

> FinCEN's guidance explains that administrators or exchangers of virtual currencies have registration requirements and a broad range of AML program, recordkeeping, and reporting responsibilities. Those offering virtual currencies must comply with these regulatory requirements. . . .
>
> Those who are intermediaries in the transfer of virtual currencies from one person to another person, or to another location, are money transmitters that must register with FinCEN as MSB [money service business], unless an exception applies. Some virtual currency exchangers have already registered with FinCEN as MSBs, though they have not necessarily identified themselves as money transmitters.[83]

---

[81] *Id.* (citing Jennifer Shasky Calvery, Director, Fin. Crimes Enforcement Network, Remarks at the Independent Armored Car Operators Association Cash in Transit Networking Conf. (May 18, 2014)).
[82] *See* Christian Catalini & Joshua S. Gans, Some Simple Economics of the Blockchain, Rotman School of Management Working Paper No. 2874598; MIT Sloan Research Paper No. 5191-16, https://ssrn.com/abstract=2874598; John M. Griffin & Amin Shams, Is Bitcoin Really Un-Tethered?, (2018), https://ssrn.com/abstract=3195066; Garrick Hileman & Michel Rauchs, 2017 Global Blockchain Benchmarking Study (2017), https://ssrn.com/abstract=3040224; MARC PILKINGTON, BLOCKCHAIN TECHNOLOGY: PRINCIPLES AND APPLICATIONS, RESEARCH HANDBOOK ON DIGITAL TRANSFORMATIONS, eds. F. Xavier Olleros and Majlinda Zhegu. (Edward Elgar, 2016), https://ssrn.com/abstract=2662660.
[83] *See* Trautman, *Virtual Currencies*, *supra* note 60 at 32 (citing *Beyond Silk Road: Potential Risks, Threats, and Promises of Virtual Currencies: Hearings Before the S. Comm. on Homeland Security & Governmental Affairs*, at 5 (statement of Jennifer Shasky Calvery, Director, Fin. Crimes Enforcement Network, U.S. Dep't. of the Treasury)).

### G. Threat of Cyber Hack

Just like: individuals;[84] corporations;[85] non-profits;[86] educational;[87] state and local governmental institutions;[88] and institutions engaged in the preservation of U.S. national security interests[89]–the universe of virtual currencies markets and currency exchangers have been subject to cyber breach.[90] As discussed more fully in our coverage of Initial Coin Offerings and The DAO funding (*see* Section VI, *Infra*), the SEC reports, "After DAO Tokens were sold, but before The DAO was able to commence funding projects, an attacker used a flaw in The DAO's code to steal approximately one-third of The DAO's assets."[91]

It is important to note that most major breaches of cryptocurrency are not due to the underlying blockchain code being exploited, but rather the supporting software which is implementing it. These breaches tend to occur at coin exchanges or currency managers.[92] Journalist Mike Orcutt reports, "hackers have stolen nearly $2 billion worth of cryptocurrency since the beginning of 2017, mostly from exchanges, and that's just what has been revealed publicly."[93] Mr. Orcutt continues:

---

[84] *See generally* Trautman, *Virtual Currencies*, *supra* note 60.

[85] *See* David D. Schein & Lawrence J. Trautman, The Dark Web & Employer Liability (unpublished manuscript), http://ssrn.com/abstract=3251479.; Lawrence J. Trautman, *How Google Perceives Customer Privacy, Cyber, E-commerce, Political and Regulatory Compliance Risks,* 10 WM. & MARY BUS. L. REV. (2018), https://ssrn.com/abstract=3067298; Lawrence J. Trautman & Peter C. Ormerod, *Corporate Directors' and Officers' Cybersecurity Standard of Care: The Yahoo Data Breach,* 66 AM. U. L. REV. 1231 (2017), http://ssrn.com/abstract=2883607; Lawrence J. Trautman & Kara Altenbaumer-Price, *The Board's Responsibility for Information Technology Governance*, 29 J. MARSHALL J. COMPUTER & INFO. L. 313 (2011), http://www.ssrn.com/abstract=1947283; Lawrence J. Trautman, *E-Commerce and Electronic Payment System Risks: Lessons from PayPal,* 17 U.C. DAVIS BUS. L.J. 261 (Spring 2016), http://www.ssrn.com/abstract=2314119; Lawrence J. Trautman, *Managing Cyberthreat,* 33(2) SANTA CLARA HIGH TECH. L.J. 230 (2017), http://ssrn.com/abstract=2534119; Lawrence J. Trautman, *The Board's Responsibility for Crisis Governance,* 13 HASTINGS BUS. L.J. 275 (2017), http://ssrn.com/abstract=2623219.

[86] *See* Lawrence J. Trautman & Janet Ford, Nonprofit Governance: The Basics, 52 AKRON L. REV. (forthcoming), https://ssrn.com/abstract=3133818.

[87] *See* Lawrence J. Trautman & Peter C. Ormerod, WannaCry, Ransomware, and the Emerging Threat to Corporations, TENN. L. REV. (2019), http://ssrn.com/abstract=3238293.

[88] *Id.*

[89] *See* Lawrence J. Trautman, *Congressional Cybersecurity Oversight: Who's Who & How It Works,* 5 J. L. & CYBER WARFARE 147 (2016), http://ssrn.com/abstract=2638448; Lawrence J. Trautman, *Is Cyberattack The Next Pearl Harbor?,* 18 N.C. J. L. & TECH. 232 (2016), http://ssrn.com/abstract=2711059; Lawrence J. Trautman, *Cybersecurity: What About U.S. Policy?,* 2015 U. ILL. J.L. TECH. & POL'Y 341 (2015), http://ssrn.com/abstract=2548561.

[90] *See, e.g.*, Paul Vigna, *Exchanges Pose Cryptocurrency Risk*, WALL ST. J., Mar. 5, 2018, at B8.

[91] *See* Report of Investigation Pursuant to Section 21(a) of the Securities Exchange Act of 1934: The DAO, SEC Release No. 81207, at 1 (July 25, 2017), https://www.sec.gov/litigation/investreport/34-81207.pdf.

[92] *Id.*

[93] *See* Mike Orcutt, *Once Hailed as Unhackable, Blockchains Are Now Getting Hacked,* MIT TECH. REV. (online) (Feb. 19, 2019), https://www.technologyreview.com/s/612974/once-hailed-as-unhackable-blockchains-are-now-getting-hacked/.

Toward the middle of 2018, attackers began springing 51% attacks on a series of relatively small, lightly traded coins including Verge, Monacoin, and Bitcoin Gold, stealing an estimated $20 million in total. In the fall, hackers stole around $100,000 using a series of attacks on a currency called Vertcoin. The hit against Ethereum Classic, which needed more than $1 million, was the first against a top-20 currency.
David Vorick, cofounder of the blockchain-based file storage platform Sia, predicts that 51% attacks will continue to grow in frequency and severity, and that exchanges will take the brunt of the damage caused by double-spends.[94]

### H. Terrorism Use

On March 20, 2018, the U.S. House Terrorism and Illicit Finance Subcommittee conducted hearings on the topic "Exploring the Financial Nexus of Terrorism, Drug Trafficking, and Organized Crime," observing that Transnational Criminal Organizations (TCOs):

> have an estimated value of $3.6 to $4.8 trillion, or seven percent of global Gross Domestic Product, and result in $130 billion in lost revenue annually to the private sector. TCOs should be regarded as a national security threat that is undermining U.S. government efforts to combat illegal drugs, arms, human trafficking, terrorism, and other crimes to include money laundering, cybercrimes, fraud, and corruption. Given the profit potential, terrorist and insurgent groups have been steadily incorporating criminal activities into their business models, thus blurring the line between TCOs and terrorist organizations.[95]

Cryptocurrency is the preferred currency for not only cyber criminals, as seen in the thousands of ransomware attacks,[96] but also for criminals operating through the internet. TOR, a (mostly) anonymous private network sometimes referred to as the dark net, leverages Bitcoin and other major cryptocurrencies to provide a level of trusted assurance and anonymity amongst parties who buy, trade, and sell everything from drugs to weapons to people. It is important to note that not everything described is illicit on the dark net or TOR as it can provide access to things that some countries lack, such as free speech,[97] or simply can provide anonymity and access to more goods.

---

[94] *Id.*
[95] Exploring the Financial Nexus of Terrorism, Drug Trafficking, and Organized Crime, Staff memorandum of the H. Terrorism and Illicit Fin. Subcomm., Mar. 15, 2018, 115th Cong. (2018).
[96] *See* Brian Krebs, *'Petya' Ransomware Outbreak Goes Global,* Krebs on Security (June 17, 2017); Brian Krebs, *After AlphaBay's Demise, Customers Flocked to Dark Market Run By Dutch Police,* Krebs on Security (July 17, 2017).
[97] *See* Robert W. Gehl, *Power/Freedom on the Dark Web: A Digital Ethnography of the Dark Web Social Network,* New Media & Society (2014), DOI: 10:1177/1461444814554900

## VI. INTERNET SCAMS ABOUND

Virtual currencies such as Bitcoin, Etherium, Ripple, et. al., along with other new applications of blockchain create a seemingly never-ending challenge for law and regulation to keep pace.[98]

With rapid advances in technology, global regulatory and law enforcement bodies encounter novel and difficult enforcement challenges. In the United States, regulators such as the Commodity Futures Trading Commission (CFTC), Securities and Exchange Commission (SEC), and the Financial Crimes Enforcement Network (FinCEN) division of The Department of the Treasury must respond to these new demands.

Do we want to discuss other country's attempts to regulate? Such as china who has banned cryptocurrency and is looking to issue their own state currency or (south) Korea who initially banned it but repealed the ban due to enforcement issues and civil push back.

### A. Securities and Exchange Commission

Cyber-securities fraud remains an unanticipated challenge to the regulation of global and U.S. securities markets, certainly "unimagined over eighty years ago by drafters of the Securities and Exchange Acts."[99] During 2018, SEC Commissioner Hester M. Peirce states, "[i]nnovation is always a challenge for regulators. We are used to the way things have been done. Our rules have grown up in response to past technologies. Figuring out whether and how they apply to new ideas is difficult."[100] In a statement dated December 11, 2017, SEC Chairman Jay Clayton observes:

---

[98] *See* Lawrence J. Trautman, *Bitcoin, Virtual Currencies and the Struggle of Law and Regulation to Keep Pace,* 102 MARQ. L. REV. 447 (2018), https://ssrn.com/abstract=3182867. *See also* Walter Keith Robinson & Joshua T. Smith, Emerging Technologies Challenging Current Legal Paradigms, 19(2) MINN. J. L. SCI. & TECH., (2018), https://ssrn.com/abstract=3220440.

[99] *See* Lawrence J. Trautman & George P. Michaely, *The SEC & The Internet: Regulating the Web of Deceit,* 68 CONSUMER FIN. L.Q. REP. 262 (2014), *citing* Securities Act of 1933 [hereinafter 1933 Act] § 2(a)(1), 15 U.S.C. § 77b(a)(1); Securities Exchange Act of 1934 [hereinafter 1934 Act] § 3(a)(10), 15 U.S.C. § 78c(a)(10), http://www.ssrn.com/abstract=1951148.

[100] Hester M. Peirce, Comm'r, U.S. Sec. and Exch. Comm'n, Beaches and Bitcoin: Remarks Before the Medici Conf., Los Angeles, Cal. (May 2, 2018), https://www.sec.gov/news/speech/speech-peirce-050218; *accord* Michael Abramowicz, *Cryptocurrency-Based Law*, 58 ARIZ. L. REV. 359, 360 (2016); Shawn Bayern, *Dynamic Common Law and Technological Change: The Classification of Bitcoin*, 71 WASH. & LEE L. REV. ONLINE 22, 23 (2014); Jerry Brito, Houman Shadab & Andrea Castillo, *Bitcoin Financial Regulation: Securities, Derivatives, Prediction Markets, and Gambling*, 16 COLUM. SCI. & TECH. L. REV. 144, 146 (2014); Chris Brummer & Yesha Yadav, *Fintech and the Innovation Trilemma*, GEO. L.J. (forthcoming 2018) (manuscript at 2), https://ssrn.com/abstract=3054770; Mark Edwin Burge, *Apple Pay, Bitcoin, and Consumers: The ABCs of Future Public Payments Law*, 67 HASTINGS L.J. 1493, 1495 (2016); Omri Marian, *A*

> The world's social media platforms and financial markets are abuzz about cryptocurrencies and "initial coin offerings" (ICOs).
>
> . . . .
>
> A number of concerns have been raised regarding the cryptocurrency and ICO markets, including that, as they are currently operating, there is substantially less investor protection than in our traditional securities markets, with correspondingly greater opportunities for fraud and manipulation.
>
> . . . [T]o date no initial coin offerings have been registered with the SEC. The SEC also has not to date approved for listing and trading any exchange-traded products (such as EFTs) holding cryptocurrencies or other assets related to cryptocurrencies.[101]

Representative of recent SEC enforcement actions regarding virtual currencies or initial coin offerings is the settlement announced on December 12, 2018, wherein:

> Two former executives behind an allegedly fraudulent initial coin offering (ICO) that was stopped by the Securities and Exchange Commission earlier this year have been ordered in federal court to pay nearly $2.7 million and prohibited from serving as officers or directors of public companies or participating in future offerings of digital securities.
>
> AriseBank's then-CEO Jared Rice Sr. and then-COO Stanley Ford were accused of offering and selling unregistered investments in their purported "AriseCoin" cryptocurrency by depicting AriseBank as a first-of-its-kind decentralized bank offering a variety of services to retail investors.
>
> "Rice and Ford lied to AriseBank's investors by pitching the company as a first-of-its kind decentralized bank offering its own cryptocurrency for customer products and services," said Shamoil T. Shipchandler, Director of the SEC's Fort Worth Regional Office. "The officer-and-director bar and digital securities offering bar will prevent Rice and Ford from engaging in another cryptoasset-based fraud."

---

*Conceptual Framework for the Regulation of Cryptocurrencies*, 82 U. CHI. L. REV. ONLINE 53, 57 (2015); Hossein Nabilou & André Prüm, *Ignorance, Debt and Cryptocurrencies: The Old and the New in the Law and Economics of Concurrent Currencies*, J. FIN. REG. (forthcoming) (manuscript at 2), https://ssrn.com/abstract=3121918; Seth C. Oranburg, *Hyperfunding: Regulating Financial Innovations*, 89 U. COLO. L. REV. 1033, 1034 (2018); Carla L. Reyes, *Moving Beyond Bitcoin to an Endogenous Theory of Decentralized Ledger Technology Regulation: An Initial Proposal*, 61 VILL. L. REV. 191, 193 (2016); Carla L. Reyes, *Conceptualizing Cryptolaw*, 96 NEB. L. REV. 384, 399 (2017); Usha Rodrigues, *Law and the Blockchain*, 104 IOWA L. REV. (forthcoming 2018) (manuscript at 50), https://ssrn.com/abstract=3127782; Scott Shackelford & Steve Myers, *Block-by-Block: Leveraging the Power of Blockchain Technology to Build Trust and Promote Cyber Peace*, 19 YALE J. L. & TECH. 335, 366 (2017).

[101] *See* Jay Clayton, *Statement on Cryptocurrencies and Initial Coin Offerings*, U.S. SEC. EXCH. COMM'N (Dec. 11, 2017), https://www.sec.gov/news/public-statement/statement-clayton-2017-12-11.

> To settle the SEC's charges, Rice and Ford agreed to be held jointly and severally liable for $2,259,543 in disgorgement plus $68,423 in prejudgment interest, and each must pay a $184,767 penalty. They also agreed to lifetime bars from serving as officers and directors of public companies and participating in digital securities offerings, and permanent prohibitions against violating the antifraud and registration provisions of the federal securities laws. . . .[102]

In terms of SEC focus on distributed ledger technology, digital assets and Initial Coin Offerings, SEC Chairman Jay Clayton states that an:

> area where the Commission and staff have spent a significant amount of time relates to distributed ledger technology, digital assets and initial coin offerings (ICOs). I expect that trend will continue in 2019. A number of concerns have been raised regarding the digital assets and ICO markets, including that, as they are currently operating, there is substantially less investor protection than in the traditional equities and fixed income markets, with correspondingly greater opportunities for fraud and manipulation.
> I believe that ICOs can be effective ways for entrepreneurs and others to raise capital. However, the novel technological nature of an ICO does not change the fundamental point that, when a security is being offered, our securities laws must be followed.
> In an effort to centralize and better coordinate the staff's work on these important issues, the SEC recently announced the formation of a new Strategic Hub for Innovation and Financial Technology ("FinHub") within the agency. Staffed by representatives from across the Commission, the FinHub serves as a public resource for fintech-related issues at the SEC. As the FinHub and our other activities demonstrate, our door remains open to those who seek to innovate and raise capital in accordance with the law.[103]

### B. Commodity Futures Trading Commission

The CFTC claims regulatory authority over digital assets (virtual currencies) under a theory that the CFTC's authority over futures and other derivatives extends to virtual currencies as commodities as defined in Section 1a(9) of the Commodity Exchange Act.[104] Based on the important characteristics of digital assets such as bitcoin, these assets, like commodities, "are units of commerce that are interchangeable, traded in markets where customers are not readily identifiable, and are immediately marketable at quoted prices. Further, like gold bullion and other commodities, bitcoin come into supply only when they are mined or extracted and are a limited resource."[105]

---

[102] Press Release, SEC, *Executives Settle ICO Scam Charges* (Dec. 12, 2018), https://www.sec.gov/news/press-release/2018-280.
[103] Jay Clayton, Chairman, U.S. Securities and Exchange Commission, SEC Rulemaking Over the Past Year, the Road Ahead and Challenges Posed by Brexit, LIBOR Transition and Cybersecurity Risks (Dec. 6, 2018), https://www.sec.gov/news/speech/speech-clayton-120618.
[104] *See* 7 U.S.C. § 1a(9) (2012); *see also* Coinflip, Inc., CFTC No. 15-29 (Sept. 17, 2015); BFXNA Inc., CFTC No. 16-19 (June 2, 2016).

In a joint statement about virtual currency enforcement actions by SEC and CFTC Enforcement Directors, Stephanie Avakian and Steven Peikin of the SEC and CFTC Enforcement Director James McDonald warn:

> When market participants engage in fraud under the guise of offering digital instruments–whether characterized as virtual currencies, coins, tokens, or the like–the SEC and the CFTC will look beyond form, examine the substance of the activity and prosecute violations of the federal securities and commodities laws. The Divisions of Enforcement for the SEC and CFTC will continue to address violations and bring actions to stop and prevent fraud in the offer and sale of digital instruments.[106]

## VII. USES FOR BLOCKCHAIN

A comprehensive discussion of the numerous potential identified use cases for blockchain technologies to date far exceeds the scope of this article. While not exhaustive, we have identified the following promising potential blockchain applications: accounting and auditing; agriculture; anti-money laundering; artificial intelligence; bills of lading; business supply chains; carbon markets; commercial real estate; commodity platform; copyrights; corporate governance; creative and artistic industries; deceptive counterfeit prevention; economic planning; education; elections; entrepreneurship and innovation; fiat money; financial services and capital markets; fractional ownership; governance; healthcare; insurance; internet of things (IoT); knowledge management; law enforcement; legal practice; marketing; privacy; promotion of world peace; property law; renewable energies; satellite navigation; securities clearing and settlement; smart cities; smart contracts; state and municipal operating efficiencies; and tax calculation and compliance. Space limitation precludes our coverage of little more than brief mention of these topics. However, a hopefully useful overview follows.

---

[105] *See* Winklevoss Bitcoin Tr., Amendment No. 9 to Form S-1 (Form S-1), at 4, F8 (Feb. 8, 2017), https://www.sec.gov/Archives/edgar/data/1579346/000119312517034708/d296375ds1a.htm.
[106] Press Release, Joint Statement by SEC and CFTC Enforcement Directors Regarding Virtual Currency Enforcement Actions: SEC Co-Enforcement Directors Stephanie Avakian and Steven Peikin and CFTC Enforcement Director James McDonald (Jan. 19, 2018), https://www.sec.gov/news/public-statement/joint-statement-sec-and-cftc-enforcement-directors.

### A. Accounting and Auditing

Suggested or documented applications for blockchain in the profession and practice of accounting and auditing are numerous.[107]

### B. Agriculture

Law firm Michael, Best & Friedrich report, "a recent surge in large food and retail firms looking to invest in the [blockchain] technology as a means to increase supply chain transparency in the food system."[108] Examples of significant industry participants include, "IBM partner[ing] with Nestle, Unilever, Tyson Foods, Walmart, and other food companies to use blockchain to increase traceability and tracking of certain products."[109] The largest U.S. retailer and employer, "Walmart is now requiring direct suppliers of lettuce, spinach, and other greens as well as farmers, logistics firms, and other supply partners to join the big-box store's food-tracking blocking starting in 2019."[110] Other applications of blockchain to food and agriculture are noted.[111]

### C. Anti-money Laundering

The use of virtual currencies by transnational crime syndicates and other corrupt actors has presented global law enforcement agencies with new and difficult challenges.[112] A new literature about virtual currency threats and blockchain applications related to anti-money laundering efforts is emerging.[113]

---

[107] *See* Volodymyr Babich & Gilles Hilary, Blockchain and Other Distributed Ledger Technologies in Operations, (unpublished ms.), https://ssrn.com/abstract=3232977; Sean S. Cao, Lin William Cong & Baozhong Yang, Auditing and Blockchains: Pricing, Misstatements, and Regulation, (unpublished ms.), https://ssrn.com/abstract=3248002; Maria Karajovic, Henry Kim & Marek Laskowski, Thinking Outside the Block: Projected Phases of Blockchain Integration in the Accounting Industry (2017), https://ssrn.com/abstract=2984126; Chandra Shekar Mylaavaram, R. Kumaran & R. K. Mishra, Blockchain Technology - An Exploratory Study on Its Applications, THE MGMT. ACCT. (June 2018), https://ssrn.com/abstract=3194522; Daniel E. O'Leary, Configuring Blockchain Architectures for Transaction Information in Blockchain Consortiums: The Case of Accounting and Supply Chain Systems (unpublished ms.), https://ssrn.com/abstract=3102671; Reinhard Schrank, Audit Quality, Legal Liability, and the Audit Market Under Risk Aversion (unpublished ms.), https://ssrn.com/abstract=3258555; Ting Yu, Zhiwei Lin & Qingliang Tang, Blockchain: Introduction and Application in Financial Accounting (unpublished ms.), https://ssrn.com/abstract=3258504.
[108] *See* Emily R. Lyons, David A. Crass, Cheryl I. Aaron & Sarah C. Helton, What Blockchain Means for the Agriculture and Food Industries, Michael Best & Friedrich LLP (Dec. 26, 2018), https://www.michaelbest.com/Newsroom/192905/What-Blockchain-Means-for-the-Agriculture-and-Food-Industries.
[109] *Id.*
[110] *Id.*
[111] *See* Henry Kim & Marek Laskowski, Agriculture on the Blockchain: Sustainable Solutions for Food, Farmers, and Financing (2017), https://ssrn.com/abstract=3028164.

### D. Artificial Intelligence

Examples of potential uses for blockchain technology with artificial intelligence (AI) are now appearing in the literature.[114]

### E. Bills of Lading

According to Elson Ong, "A bill of lading operates as a document of title if it enables the consignee to take delivery of the goods at their destination or to dispose of them by the endorsement and delivery of the bill of lading."[115] Other scholars have discussed the application of blockchain technology to bills of lading.[116]

### F. Business Supply Chains

Business supply chains appear to be an area having unusually robust promise for successful blockchain application.[117]

---

[112] *See* Sean Foley, Jonathan R. Karlsen & Talis J. Putnins, Sex, Drugs, and Bitcoin: How Much Illegal Activity Is Financed Through Cryptocurrencies?, REV. FIN. STUD. (Forthcoming), https://ssrn.com/abstract=3102645.

[113] *See* Ross Anderson, Ilia Shumailov, Mansoor Ahmed & Alessandro Rietmann, Bitcoin Redux, at 6.5 (2018); Ehi Esoimeme, A Critical Analysis of the Anti-Money Laundering Measures Adopted by BitGold Inc. (2015), https://ssrn.com/abstract=2604721; Ehi Esoimeme, Balancing Anti-Money Laundering Measures and Financial Inclusion: The Example of the United Kingdom and Nigeria, J. MONEY LAUNDERING CONTROL, (Forthcoming), https://ssrn.com/abstract=3172081.

[114] *See* Bo Xing, & Tshilidzi Marwala, The Synergy of Blockchain and Artificial Intelligence (2018), https://ssrn.com/abstract=3225357. *See also* William Magnuson, Artificial Financial Intelligence (unpublished ms.) (2019).

[115] *See* Elson Ong, *Call a Bill a Bill: The Star Quest*, 23 J. INT'L MARITIME L. 328, 334 (2017), https://ssrn.com/abstract=3123753.

[116] *See* Elson Ong, Blockchain Bills of Lading. NUS Law Working Paper No. 2018/020 (2018), https://ssrn.com/abstract=3225520.

[117] *See* Volodymyr Babich & Gilles Hilary, Distributed Ledgers and Operations: What Operations Management Researchers Should Know About Blockchain Technology (2018). Forthcoming in Manufacturing & Service Operations Management; Georgetown McDonough School of Business Research Paper No. 3131250. https://ssrn.com/abstract=3131250; Chris Berg, Sinclair Davidson & Jason Potts, Outsourcing Vertical Integration: Distributed Ledgers and the V-Form Organisation (2018), https://ssrn.com/abstract=3300506; Bhavya Bhandari, Supply Chain Management, Blockchains and Smart Contracts (2018), https://ssrn.com/abstract=3204297; Jiri Chod, Nikolaos Trichakis, Gerry Tsoukalas, Henry Aspegren & Mark Weber, Blockchain and the Value of Operational Transparency for Supply Chain Finance 2018). Mack Institute for Innovation Management, Working Paper Series, https://ssrn.com/abstract=3078945 (providing signals enhancing finance); Henry Kim & Marek Laskowski, Towards an Ontology-Driven Blockchain

### G. Carbon Markets

Several scholars have written about the application of blockchain technology to carbon markets.[118] See also our separate discussion regarding renewable energy (*Infra.*).

### H. Commercial Real Estate

A literature review reveals a modest trail of scholarship to date illustrating blockchain applications to commercial real estate.[119] See also coverage of property law and real estate (*Infra*).

### I. Commodity Platforms

Commodity trade finance platform proposals using distributed ledger technology has been noted.[120]

### J. Copyrights

A number of scholars have written about the application of blockchain technology to copyright.[121]

---

Design for Supply Chain Provenance (2016), https://ssrn.com/abstract=2828369; Adam J. Sulkowski, Blockchain, Law, and Business Supply Chains: The Need for Governance and Legal Frameworks to Achieve Sustainability (2018), https://ssrn.com/abstract=3205452.

[118] *See* Adrian Jackson, Ashley Lloyd, Justin Macinante & Markus Hüwener, Networked Carbon Markets: Permissionless Innovation with Distributed Ledgers?, (2018). Edinburgh Sch. L. Res. Paper No. 2018/07, https://ssrn.com/abstract=3138478; Robert Leonhard, Developing the Crypto Carbon Credit on Ethereum's Blockchain (2017), https://ssrn.com/abstract=3000472; Robert Leonhard, Forget Paris: Building a Carbon Market in the U.S. Using Blockchain-Based Smart Contracts (2017), https://ssrn.com/abstract=3082450; Justin Macinante, A Conceptual Model for Networking of Carbon Markets on Distributed Ledger Technology Architecture. Edinburgh Sch. L. Res. Paper No. 09/2017, https://ssrn.com/abstract=2948580.

[119] *See* Hitesh Malviya, Blockchain for Commercial Real Estate (2017), https://ssrn.com/abstract=2922695; Sergio Nasarre-Aznar, Collaborative Housing and Blockchain, 66(2) ADMINISTRATION 59 (2018), https://ssrn.com/abstract=3189050.

[120] *See* Jianfu Wang, Commodity Trade Finance Platform Using Distributed Ledger Technology: Token Economics in a Closed Ecosystem Using Agent-Based Modeling (2018), https://ssrn.com/abstract=3152093.

[121] *See* Annabel Tresise, Jake Goldenfein & Dan Hunter, *What Blockchain Can and Can't Do for Copyright*, 28 AUSTL. INTELL. PROP. J. 144 (2018), https://ssrn.com/abstract=3227381; Jake Goldenfein & Dan Hunter, Blockchains, Orphan Works, and the Public Domain, 41 COLUM. J.L. & ARTS (2017), https://ssrn.com/abstract=3083153; Nick Vogel, The Great Decentralization: How Web 3.0 Will Weaken Copyrights, 15 J. MARSHALL REV. INTELL. PROP. L. 136 (2015).

### K. Corporate Governance

Directors are responsible for the governance of a corporation.[122] Numerous scholars have found blockchain applications to be applicable to the governance of corporations.[123]

### L. Creative and Artistic Industries

Outside the specific category of copyright, a number of scholars have written about the potential value of blockchain to a broad category of endeavors we are describing as the creative and artistic industries.[124]

### M. Deceptive Counterfeit Prevention

The potential for deployment of blockchain technology in the prevention of deceptive counterfeit goods is noted.[125]

---

[122] *See* Lawrence J. Trautman & Peter C. Ormerod, WannaCry, Ransomware, and the Emerging Threat to Corporations, TENN. L. REV. (2019), http://ssrn.com/abstract=3238293.

[123] *See* Paul H. Edelman, Wei Jiang & Randall S. Thomas, Will Tenure Voting Give Corporate Managers Lifetime Tenure?, Vanderbilt Law Research Paper No. 18-04; European Corporate Governance Institute (ECGI) - Law Working Paper No. 384/2018. (2018), https://ssrn.com/abstract=3107225; Mark Fenwick & Erik P.M. Vermeulen, Technology and Corporate Governance: Blockchain, Crypto, and Artificial Intelligence (2018). Lex Research Topics in Corporate Law & Economics Working Paper No. 2018-7; Eur. Corp. Governance Institute (ECGI) - Law Working Paper No. 424/2018. https://ssrn.com/abstract=3263222; Mark Fenwick, Joseph A. McCahery & Erik P.M. Vermeulen, The End of 'Corporate' Governance: Hello 'Platform' Governance, Lex Research Topics in Corporate Law & Economics Working Paper No. 2018-5; European Corporate Governance Institute (ECGI) - Law Working Paper No. 430/2018, https://ssrn.com/abstract=3232663; Robert Leonhard, Corporate Governance on Ethereum's Blockchain (2017), https://ssrn.com/abstract=2977522; Christoph Van der Elst & Anne Lafarre, Blockchain and Smart Contracting for the Shareholder Community. European Corporate Governance Institute (ECGI) - Law Working Paper No. 412/2018. https://ssrn.com/abstract=3219146; Christoph Van der Elst & Anne Lafarre, Bringing the AGM to the 21st Century: Blockchain and Smart Contracting Tech for Shareholder Involvement (June 2017). European Corporate Governance Institute (ECGI) - Law Working Paper No. 358/2017, https://ssrn.com/abstract=2992804.

[124] *See* Jason Potts & Ellie Rennie, Blockchains and Creative Industries (2017), https://ssrn.com/abstract=3072129; Stan Sater, Tokenize the Musician, 21 Tulane J. Tech. & Intell. Prop. (2018), https://ssrn.com/abstract=3160798 (music recording and performance economics); Kelly Snook & Jason Potts, Concordia – A New Future Economy of Music. For International Cultural Economics Conference, 27-29 June, RMIT University, Melbourne Australia, https://ssrn.com/abstract=3174172; Bo Xing, Creativity and Artificial Intelligence: A Digital Art Perspective (2018), https://ssrn.com/abstract=3225323 (digital art).

[125] *See* Hubert Pun, Jayashankar M. Swaminathan & Pengwen Hou, Blockchain Adoption for Combating Deceptive Counterfeits (2018). Kenan Institute of Private Enterprise Research Paper No. 18-18., https://ssrn.com/abstract=3223656.

### N. Economic Planning

Several scholars have posited the scenario of an entire planned economic system through blockchain networks.[126]

### O. Education

A number of scholars have explored the use of blockchain technology in education, concluding that these applications are presently only in an infancy state, and offering suggestions for additional research.[127]

### P. Elections

A few scholars have written about the application of blockchain technology to elections.[128]

### Q. Entrepreneurship and Innovation

Professors Chris Berg, Sinclair Davidson, and Jason Potts contend "that the institutional innovation of blockchain engenders a new post-industrial economic era that requires new policy rules."[129] This paper seeks to explain why this change will occur, and to explore a new framework for economic policy adapted to economic infrastructure built on distributed ledgers.

### R. Extractive Industries

The benefits of blockchain technology to the extractive industries has been demonstrated by several scholars.[130]

---

[126] *See* Kartik Hegadekatti & Yatish S. G., The Programmable Economy: Envisaging an Entire Planned Economic System as a Single Computer Through Blockchain Networks, (2017), https://ssrn.com/abstract=2943227.

[127] *See* Wiebke Lévy, Jutta Stumpf-Wollersheim & Isabell M., Welpe, Disrupting Education Through Blockchain-Based Education Technology?, (2018), https://ssrn.com/abstract=3210487.

[128] *See* Usman W. Chohan, Blockchain Enhancing Political Accountability? Sierra Leone 2018 Case (2018), https://ssrn.com/abstract=3147006; Kartik Hegadekatti, Analysis of Present Day Election Processes vis-à-vis Elections Through Blockchain Technology (2017), https://ssrn.com/abstract=2904868; Samuel Martin, Blockchain as a Solution to the United States' Voter Turnout Issue (2018), https://ssrn.com/abstract=3177523.

[129] *See* Chris Berg, Sinclair Davidson & Jason Potts, *Capitalism After Satoshi: Blockchains, Dehierarchicalisation, Innovation Policy, and the Regulatory State*, J. ENTREPRENEURSHIP & PUB. POL'Y (2018), https://ssrn.com/abstract=3299776; Jon M. Truby, *FinTech and the City: Sandbox 2.0 Policy and Regulatory Reform Proposals,* INT'L REV. L. COMP. & TECH., https://ssrn.com/abstract=3299300.

[130] *See* Usman W. Chohan, Blockchain and the Extractive Industries: Cobalt Case Study (2018),

### S. Fiat Money

A number of authors explore the history and definition of money and whether crypto currencies may replace fiat currencies.[131]

### T. Financial Services and Capital Markets

Much has been written about the application of blockchain technology to financial services and capital markets. Unfortunately, space limits our footnote coverage to just a small portion of the body literature.[132]

---

https://ssrn.com/abstract=3138271; Usman W. Chohan, Blockchain and the Extractive Industries #2: Diamonds Case Study (2018), https://ssrn.com/abstract=3141883; Ushnish Sengupta, & Henry Kim, How Indigenous Entrepreneurs Can Use Trustless Technology to Rebuild Trust: A Case for Business Process Transformation in Natural Resources Development Using Blockchain (2018), https://ssrn.com/abstract=3246385.

[131] *See* Richard Senner & Didier Sornette, The Holy Grail of Crypto Currencies: Ready to Replace Fiat Money?, (Forthcoming) J. ECON. ISSUES, https://ssrn.com/abstract=3192924.

[132] *See* Lawrence J. Trautman, *Is Disruptive Blockchain Technology the Future of Financial Services?*, CONSUMER FIN. L.Q. REP. 232, 234 (2016). *See also* Jun Aoyagi, Daisuke Adachi, Economic Implications of Blockchain Platforms (2018), https://ssrn.com/abstract=3132235; Catherine Martin Christopher, The Bridging Model: Exploring the Roles of Trust and Enforcement in Banking, Bitcoin, and the Blockchain, 17 NEVADA L.J. 1 (2016), https://ssrn.com/abstract=2851492; David Lee Kuo Chuen, Decentralization and Distributed Innovation: Fintech, Bitcoin and ICO's (2017), https://ssrn.com/abstract=3107659; Ryan Clements, Evaluating the Costs and Benefits of a Smart Contract Blockchain Framework for Credit Default Swaps, 10 WILLIAM & MARY BUS. L. REV. 3 (2019), https://ssrn.com/abstract=3197706; Shaen Corbet, Charles James Larkin, Brian M. Lucey, Andrew Meegan Larisa Yarovaya, Cryptocurrency Reaction to FOMC Announcements: Evidence of Heterogeneity Based on Blockchain Stack Position (2017), https://ssrn.com/abstract=3073727; Benjamin Geva, Banking in the Digital Age - Who is Afraid of Payment Disintermediation? (2018). European Banking Institute Working Paper Series 2018 - no. 23., https://ssrn.com/abstract=3153760; Giuseppe Giudici, Legal Problems of the Blockchain: A Capital Markets Perspective (2018), https://ssrn.com/abstract=3240273; Peter Gomber, Robert J. Kauffman, Chris Parker & Bruce Weber, On the Fintech Revolution: Interpreting the Forces of Innovation, Disruption and Transformation in Financial Services, 35(1) J. MGMT. INFO. SYS. 220 (2018), https://ssrn.com/abstract=3190052; Adam Hayes, Decentralized Banking: Monetary Technocracy in the Digital Age, Tenth Mediterranean Conference on Information Systems (MCIS), Paphos, Cyprus, September 2016, https://ssrn.com/abstract=2807476; Gur Huberman, Jacob Leshno & Ciamac C. Moallemi, An Economic Analysis of the Bitcoin Payment System (2018). Columbia Business School Research Paper No. 17-92., https://ssrn.com/abstract=3025604; Sarah Jane Hughes, *Permission-less Blockchain-Based Payment and Property Transactions Challenge Global Anti-Money-Laundering and Counter-Terrorism Finance Laws Enforcement Regimes* (unpublished ms.) (on file with authors); Anil Savio Kavuri & Alistair K. L. Milne, Fintech and the Future of Financial Services: What are the Research Gaps? (2018), https://ssrn.com/abstract=3215849; Christoffer Koch & Gina C. Pieters, Blockchain Technology Disrupting Traditional Records Systems (July 5, 2017). Financial Insights - Dallas Federal Reserve Bank, https://ssrn.com/abstract=2997588; Joseph Lee, Distributed Ledger Technologies (Blockchain) in Capital Markets: Risk and Governance (2018), https://ssrn.com/abstract=3180553; William J. Magnuson, Regulating Fintech, 71(4) VAND. L. REV. 1167 (2018), https://ssrn.com/abstract=3027525; William J. Magnuson, *Financial Regulation in the Bitcoin Era,* 23(2) STAN. J.L. BUS. & FIN. 159 (2018), https://ssrn.com/abstract=3148036; Daniel E. O'Leary,

### U. Fractional Ownership

Literature has recently appeared illustrating the application of blockchain technology to facilitate fractional ownership.[133] See also our discussion regarding the creative industries (*Infra*).

### V. Governance

Many applications fall into this category. One of the idealized uses of blockchain is to track, with irrefutable certainty, events such as financial transactions or health services.[134]

---

Open Information Enterprise Transactions: Business Intelligence and Wash and Spoof Transactions in Blockchain and Social Commerce (2018), https://ssrn.com/abstract=3246740; José Parra-Moyano, Tryggvi Thoroddsen & Omri Ross, Optimized and Dynamic KYC System Based on Blockchain Technology (March 14, 2018), https://ssrn.com/abstract=3248913; Max Raskin & David Yermack, Digital Currencies, Decentralized Ledgers, and the Future of Central Banking (2016). NBER Working Paper No. w22238, https://ssrn.com/abstract=2777326; Margaret Ryznar, The Future of Bitcoin Futures, __ HOUSTON L. REV. (Forthcoming), https://ssrn.com/abstract=3127327; Lawrence J. Trautman &n Oliver W. Aho), *Crowdfunding, Entrepreneurship, and Start-Up Finance,* http://ssrn.com/abstract=325138; Anne M. Tucker & Holly van den Toorn, Will Swing Pricing Save Sedentary Shareholders?, 2018(1) COLUM. BUS. L. REV. __ (2018), https://ssrn.com/abstract=3173736; Martin Walker, Bridging the Gap between Investment Banking Architecture and Distributed Ledgers (2017), https://ssrn.com/abstract=2939281.

[133] *See* Amy Whitaker & Roman Kraeussl, Democratizing Art Markets: Fractional Ownership and the Securitization of Art (2018), https://ssrn.com/abstract=3100389.

[134] *See* Marcella Atzori, Blockchain Governance and the Role of Trust Service Providers: The TrustedChain® Network (2017), https://ssrn.com/abstract=2972837; Alastair Berg, Chris Berg & Mikayla Novak, Crypto Public Choice (2018), https://ssrn.com/abstract=3236025; Balázs Bodó, João Quintais, Alexandra Giannopoulou & Valeria Ferrari, Barlaeus Dinner on Trust in Decentralized Data Infrastructures, Blockchain & Society Policy Research Lab Research Nodes 2018/2, https://ssrn.com/abstract=3243360; Nick Cowen, Markets for Rules: The Promise and Peril of Blockchain Distributed Governance, https://ssrn.com/abstract=3223728; Primavera De Filippi & Samer Hassan, Blockchain Technology as a Regulatory Technology: From Code is Law to Law is Code (December 5, 2016). Primavera De Filippi & Samer Hassan, Blockchain Technology as a Regulatory Technology: From Code is Law to Law is Code. 21(12) First Monday, special issue on 'Reclaiming the Internet with distributed architectures (2016), https://ssrn.com/abstract=3097430; Samer Hassan & Primavera De Filippi, The Expansion of Algorithmic Governance: From Code Is Law to Law Is Code, In Hassan, S. & De Filippi, P. (2017). The Expansion of Algorithmic Governance: From Code is Law to Law is Code. Field Actions Science Reports: The Journal of Field Actions. Special issue 17: Artificial Intelligence and Robotics in the City. Open Edition Journals, https://ssrn.com/abstract=3117630; Ying-Ying Hsieh, Jean-Philippe Vergne & Sha Wang, The Internal and External Governance of Blockchain-Based Organizations: Evidence from Cryptocurrencies, (Forthcoming) in Campbell-Verduyn M (ed.), Bitcoin and Beyond: Blockchains and Global Governance. RIPE/Routledge Series in Global Political Economy, https://ssrn.com/abstract=2966973; Brendan Markey-Towler, Anarchy, Blockchain and Utopia: A Theory of Political-Socioeconomic Systems Organised using Blockchain (2018), https://ssrn.com/abstract=3095343; Gianluca Miscione, Rafael Ziolkowski, Liudmila Zavolokina &

### W. Healthcare

Healthcare in a number of various forms appears to be a category presenting substantial potential benefit from blockchain technology.[135] However, rapid technological advances may bring serious new privacy concerns.[136]

### X. Insurance

The insurance industry has also been identified by scholars as a prime candidate for use of artificial intelligence and blockchain technology.[137]

---

Gerhard Schwabe, Tribal Governance: The Business of Blockchain Authentication, Prepared for the Hawaii International Conference on System Sciences (HICSS) (2018), https://ssrn.com/abstract=3037853; Julia M. Puaschunder, On Artificial Intelligence's Razor's Edge: On the Future of Democracy and Society in the Artificial Age (2018), https://ssrn.com/abstract=3297348; David Rozas, Antonio Tenorio-Fornés, Silvia Díaz-Molina & Samer Hassan, When Ostrom Meets Blockchain: Exploring the Potentials of Blockchain for Commons Governance (2018), https://ssrn.com/abstract=3272329; Gili Vidan & Vili Lehdonvirta, Mine the Gap: Bitcoin and the Maintenance of Trustlessness, NEW MEDIA & SOCIETY, (Forthcoming), https://ssrn.com/abstract=3225236; Karen Yeung, Regulation by Blockchain: The Emerging Battle for Supremacy between the Code of Law and Code as Law, MODERN L. REV. (Forthcoming), https://ssrn.com/abstract=3206546.

[135] *See* William J. Blackford, Hashing it Out: Blockchain as a Solution for Medicare Improper Payments, 5 BELMONT L. REV. 219 (2018), https://ssrn.com/abstract=3240973; Thomas Heston, Why Blockchain Technology Is Important for Healthcare Professionals (2017), https://ssrn.com/abstract=3006389; Thomas Heston, A Case Study in Blockchain Healthcare Innovation (2017). Authorea Working Paper No AUTHOREA_213011_3643634, https://ssrn.com/abstract=3077455; Marc Pilkington, Can Blockchain Improve Healthcare Management? Consumer Medical Electronics and the IoMT (2017), https://ssrn.com/abstract=3025393; Ana Santos Rutschman, Healthcare Blockchain Infrastructure: A Comparative Approach (July 2, 2018). Saint Louis U. Legal Studies Research Paper No. 2018-4., https://ssrn.com/abstract=3217297; Mirko Schedlbauer & Kerstin Wagner, Blockchain Beyond Digital Currencies – A Structured Literature Review on Blockchain Applications (2018), *citing* A. Roehrs, C.A. da Costa & R. d. R. Righi, OmniPHR: A Distributed Architectural Model to Integrate Personal Health Records, J. BIOMEDICAL INFOR. 71, 70 (DOI:10.1016/j.jbi.2017.05.012), https://ssrn.com/abstract=3298435; Scott J. Shackelford, Michael Mattioli, Steven Myers, Austin E. Brady, Ruihan Wang & Stephanie Wong, Securing the Internet of Healthcare, MINN. J. L., SCI. & TECH. (2018), https://ssrn.com/abstract=3128683.

[136] *See* Anthony W. Orlando & Arnold J. Rosoff, *The New Privacy Crisis: What's Health Got to Do With It?,* 132(2) AM. J. MEDICINE 127 (Feb. 2019), https://www.amjmed.com/article/S0002-9343(18)31025-8/fulltext; Anthony W. Orlando & Arnold J. Rosoff, *Fast-Forward to the Frightening Future: How the 21st Century Cures Act Accelerates Technological Innovation… at Unknown Risk to Us All,* 44 AM. J. L. & MEDICINE 253 (2018), https://journals.sagepub.com/doi/abs/10.1177/0098858818789425.

[137] *See* Angelo Borselli, Insurance by Algorithm, Eur. INS. L. REV. (2018), https://ssrn.com/abstract=3284437; Shaen Corbet & Constantin Gurdgiev, Financial Digital Disruptors and Cyber-Security Risks: Paired and Systemic, 1(2) J. TERRORISM & CYBER INS. (2017), https://ssrn.com/abstract=2892842.

### Y.  Internet of Things (IoT)

The literature suggests that the Internet of Things (IoT) may benefit from use of blockchain technology.[138]

### Z.  Law Enforcement

Professor Karen Yeong describes the benefits of how automated enforcement via distributed ledger systems might work.[139]

### AA.  Legal Practice

In addition to our discussion about smart contracts, the blockchain will likely impact the practice of law in numerous ways.[140]

---

[138] *See* Arushi Arora & Sumit Kumar Yadav, Block Chain Based Security Mechanism for Internet of Vehicles (IoV), Proceedings of 3rd International Conference on Internet of Things and Connected Technologies (ICIoTCT), 2018 held at Malaviya National Institute of Technology, Jaipur (India) on March 26-27, 2018, https://ssrn.com/abstract=3166721; Rejwan Bin Sulaiman, Applications of Block-Chain Technology and Related Security Threats (2018), https://ssrn.com/abstract=3205732; Bjorn Lundqvist, Portability in Datasets under Intellectual Property, Competition Law, and Blockchain (2018). Faculty of Law, Stockholm University Research Paper No. 62, https://ssrn.com/abstract=3278580; Lee W. McKnight, Richie Etwaru & Yihan Yu, Commodifying Trust: Trusted Commerce Policy Intersecting Blockchain and Internet of Things (2017), https://ssrn.com/abstract=2944466; Scott J. Shackelford, Governing the Internet of Everything: Applying the IAD and GKC Frameworks to Improve the Security and Privacy of Things (2018). Kelley School of Business Research Paper No. 18-86., https://ssrn.com/abstract=3266188; Scott J. Shackelford, Smart Factories, Dumb Policy?: Managing Cybersecurity and Data Privacy Risks in the Industrial Internet of Things (2018). Kelley School of Business Research Paper No. 18-80, https://ssrn.com/abstract=3252498; Lawrence J. Trautman & Peter C. Ormerod, *Industrial Cyber Vulnerabilities: Lessons from Stuxnet and the Internet of Things,* 72 U. MIAMI L. REV. 761 (2018), http://ssrn.com/abstract=2982629.

[139] *See* Karen Yeung, Blockchain, Transactional Security and the Promise of Automated Law Enforcement: The Withering of Freedom Under Law? (2017). TLI Think! Paper 58/2017; Forthcoming, iRights.Media (ed) 3TH1CS - The reinvention of ethics in the digital age (2017), iRights.Media, Berlin.; King's College London Law School Research Paper No. 2017-20, https://ssrn.com/abstract=2929266,

[140] *See* Alan Cunningham, Andrew D. James, Paul Taylor & Bruce Tether, Disruptive Technologies & Legal Service Provision in the UK: A Preliminary Study (2018), https://ssrn.com/abstract=3297074; Kate Galloway, Text to Bits: Beyond the Revolution in Law and Lawyering (2016), https://ssrn.com/abstract=2879220.

### BB. Marketing

Professors Campbell Harvey, Christine Moorman, and Marcos Castillo Toledo present a compelling argument for the first movers in applying blockchain to marketing.[141]

### CC. Privacy

The intersection of blockchain technology and privacy issues has been addressed to date by several scholars.[142]

### DD. Promotion of World Peace

Professors Scott Shackelford & Steven Myers, "examine the rise of blockchains through the lens of the literature on polycentric governance to ascertain what lessons this research holds to build trust in distributed systems and ultimately promote cyber peace."[143]

### EE. Property Law and Real Estate

Professor Katie Szilagyi discusses how blockchain has created novel and significant demands for changes in property law.[144] Real estate practitioner Avi Spielman observes, "the inevitable convergence of blockchain and real estate may be viewed as a modern-day example of the classic confrontation staged when an immovable object meets an unstoppable force."[145] Potential applications of blockchain to real estate include, "improvements in database management, information management, and efficiency—such as multiple listing services (MLS) and smart contracts (leasing)."[146]

---

[141] *See* Campbell R. Harvey, Christine Moorman & Marcos Castillo Toledo, How Blockchain Will Change Marketing as We Know It (September 29, 2018), https://ssrn.com/abstract=3257511.
[142] *See* Henry Chang, *Is Distributed Ledger Technology Built for Personal Data?,* 1(4) J. DATA PROTECTION & PRIVACY, (2018), https://ssrn.com/abstract=3137606; Jonathan Cave, Private Communication in Public Spaces: The Paradoxical Economics of Exceptional Access and E-Privacy (March 30, 2016). TPRC 44: The 44th Research Conference on Communication, Information and Internet Policy 2016, https://ssrn.com/abstract=2757699.
[143] *See* Scott J. Shackelford & Steven Myers, *Block-by-Block: Leveraging the Power of Blockchain Technology to Build Trust and Promote Cyber Peace,* 19 Yale J. L. & Tech. 334 (2017), https://ssrn.com/abstract=2874090.
[144] *See* Katie Szilagyi, A Bundle of Blockchains? Digitally Disrupting Property Law, 48 CUMB. L. REV. 9 (2018), https://ssrn.com/abstract=3161151.
[145] *See* Avi Spielman, *Blockchain and Commercial Real Estate,* SIOR 3, (2017), https://www.sior.com/docs/default-source/Thought-leadership/blockchain-and-commercial-real-estate-real-final_.pdf?sfvrsn=5cbceba8_0 (last viewed Feb. 24, 2019).

An excellent You Tube video about "Emerging Real Estate Trends in Cryptocurrency and Blockchain," is perhaps the most cogent explanation we've seen about the economics of blockchain mining.[147] Based in Montreal, Canada, CBRE's real estate professional David Cervantes, observes:

### Power Availability Dictates the Real Estate Footprint

> If we consider the global map of primary hubs for the mining process of crypto currency . . . the real driver is excess power or unused capacity at the generating source of power internationally. It is not enough to have cheap power, it must be cheap and abundant thus allowing for a marketplace to set-in.
>
> We monitor power costs across our many markets and find our clients gravitating to those regions that can offer power in the range of 2.5 to 5.5 cents US per kWh. [2016 numbers] But increasingly, we are seeing a demand for greater price certainty with guaranteed power contracts exceeding 5-year terms.
>
> When seeking the link between the demand for mining capacity and the growth of the global mining footprint, all indicators point to the contracting of power. Crypto mining operations, the end-users of real estate in this industry, are required to strike power contracts with local utility companies. Where the power utilities are ready to commit to this sector by way of predictable power contracts, we see rampant growth.
>
> In China, for years, there had been no obstacles to growth for the large crypto mining firms. [During 2016], nearly 80% of the world's crypto mining throughput was being processed in China.
>
> Utility companies, originally only a supplier of electricity to the sector have begun to partner with [blockchain] miners or have become themselves operators of crypto mines. This had led to coins acting as a unique store-of-value, not in the investment sense, but as a store of excess electrical capacity. Utility companies can tokenize their excess capacity when power is abundant and thus stabilize their profits when power is less so. . . .

### Power Cost

> To offer a window into the global marketplace for space and power, there is an emerging cost model and metric that allows market players to compare and compete internationally. The two variables are full-service power cost and facility cost. Now this is a departure from our traditional real estate inputs of dollars and square feet or square meters, but it begins belies the importance of power and cost to this sector as opposed to area. If we add the regional power cost to the facility cost and compare that figure to markets around the globe, we can quickly observe thresholds deemed appropriate by end users to pursue a deal.

---

[146] *Id.*
[147] *See* David Cervantes, "Emerging Real Estate Trends in Cryptocurrency and Blockchain," CBRE (Oct. 2, 2018), https://www.youtube.com/watch?v=BnjIV_Qcgzc (last viewed Feb. 24, 2019).

To provide an example from the wholesale colocation market:

We earlier referenced desirable power costs at 2.5 to 5.5 cents per kWh US. [2016 numbers][148]

The market rate paid for wholesale hosted transactions is 7 to 10 cents per kWh US.

The margin remaining when we deduct the local power cost from the total wholesale cost represents the rent potential of the facility. . . .

**How Regulations Act as Both a Threat and a Stabilizing Force**

Power policy and government regulation has already played a significant role in this sector and will continue to impact the attractiveness of markets. Real estate is strongly linked to power policy in this sector just as it is in other high-power consuming industries like the production of steel. . . .

As an example, the recorded demand from crypto currency miners targeting the Province of Quebec in the 5 months between October 2017 and February 2018 was 18,000 MWs. Hydro Quebec, the publicly operated utility company, called for government intervention. A moratorium was enacted, and new regulations were imposed in order to process and prioritize the demand. In order to manage the market, the government enacted both an economic development rate to be applied to firms that they deep qualified and a punitive and retroactive rate of 15 cents per KWH for any firms found to be operating without a permit. These new rules sent many end-users scrambling to determine whether they would be awarded permits and dissuaded many others from entry.

Regulation is a threat when a regional government can halt or hike prices of power. Any market which has opened its doors to new demand and who boasts alignment with government becomes a safer destination by comparison. A market which is post-regulation and/or has government alignment is even more attractive than those markets without any regulation at all.

Site selection for crypto currency is gaining sophistication as operators are expanding their criteria beyond low-cost power to now also value predictable and reliable power. In the pursuit to stem risk in an inherently risky early phase of this business, regulation, if not too heavy handed, can provide a framework for productive deal making.

---

[148] E-mail from David Cervantes, Sr. Vice. Pres., CBRE Ltd. to Lawrence J. Trautman (Mar. 4, 2019, 12:19 CST) (on file with authors) (observing "market pricing has come down since publication. Certainly $0.02 less per kwh USD in both wholesale and retail examples).

**Direct Power at High Scale**

. . . A crypto mine may not have a tremendous amount in common with a tier 3 data center, but it may begin to closely resemble a quantum computing facility, or a rendering farm used to illustrate 3 dimensional landscapes for films and video games. The growing need for compute capacity at ever-lower costs are leading end-users to explore and demand less resilient solutions. Sectors, include crypto currency, AI and others, need less of the high cost critical infrastructure such as generators and temperature control…The site selection for crypto currency will closely resemble that of large-scale quantum computing and virtual pools of 3 dimensional rendering, to name only these few.

**[Observations]…**

1. Not all regions are created equal. The crypto currency and blockchain demand will gravitate to regions that have low power cost but also have abundant capacity for which power utilities are ready to contract.
2. Power cost and facility cost are inversely proportional. Where the power cost is lowest, the margin for revenue potential of the facility is highest. If the regional power cost is too high, the revenue potential of a particular facility may be extinguished.
3. Regulation is being applied in many different ways across all our markets. We advise our clients to be aware of prospective regulation and seek alignment with government power policies be part of a solution. . . .[149]

### FF. Renewable Energies

Several scholars have explored use of blockchain to facilitate growth in the energy industry toward renewable resources.[150]

### GG. Satellite Navigation

Blockchain is . . . slowly-changing, complex systems such as satellite orbits.

Blockchain technology offers a unique solution to this problem by utilizing functions, processes, and information already incorporated into satellites while maintaining a maximum of forty-eight hours of data. The use of blockchain provides a high level of trusted positional data that can be used to predict and avoid collisions which will save billions of dollars and valuable time.[151]

---

[149] *Id.*
[150] *See* Adrienne Adjeleian, Oana Jurjica & Henry Kim, Breaking the Stagnant Spell: How Blockchain is Disrupting the Solar Energy Industry (2018), https://ssrn.com/abstract=3207104; Lauren Downes & Chris Reed, Blockchain for Governance of Sustainability Transparency in the Global Energy Value Chain, Queen Mary School of Law Legal Studies Research Paper No. 283/2018 (2018), https://ssrn.com/abstract=3236753.

### HH. Securities Clearing and Settlement

Colleen Baker has recently written an excellent account of the over-the-counter (OTC) derivative markets and clearinghouse structure and importance to global capital markets.[152] Other scholars have illustrated the potential for blockchain based applications to provide efficiencies and security in the execution of clearinghouse functions.[153]

---

[151] *See* Mason J. Molesky, Elizabeth A. Cameron, Jerry Jones IV, Michael Esposito, Liran Cohen & Chris Beauregard, *Blockchain Network for Space Object Location Gathering,* IEEE 978-1-5386-7266-2/18.

[152] *See* Colleen M. Baker, Incomplete Clearinghouse Mandates (unpublished ms.) (on file with authors). *See also* Colleen M. Baker, *Regulating the Invisible: The Case of Over-the-Counter Derivatives,* 85 NOTRE DAME L. REV. 1287 (2010); Colleen M. Baker, *The Federal Reserve as Last Resort,* 46 U. MICH J.L. REFORM 69 (2012); *When Regulators Collide: Financial Market Stability, Systemic Risk, Clearinghouses, and CDS,* 10 VA. L. & BUS. REV. 589 (2016);

[153] *See* Ferdinando M. Ametrano, Emilio Barucci, Daniele Marazzina & Stefano Zanero, Response to ESMA/2016/773: 'The Distributed Ledger Technology Applied to Securities Markets' (2016), https://ssrn.com/abstract=3265776; Emilios Avgouleas & Aggelos Kiayias, The Promise of Blockchain Technology for Global Securities and Derivatives Markets: The New Financial Ecosystem and the 'Holy Grail' of Systemic Risk Containment (2018). Edinburgh School of Law Research Paper No. 2018/43, https://ssrn.com/abstract=3297052; Jerry Brito, Houman B. Shadab & Andrea Castillo O'Sullivan, *Bitcoin Financial Regulation: Securities, Derivatives, Prediction Markets, and Gambling,* 16 COLUM. SCI. & TECH. L. REV. 144 (2014), https://ssrn.com/abstract=2423461; Joanna Caytas, *Developing Blockchain Real-Time Clearing and Settlement in the EU, U.S., and Globally,* 22 COLUM. J. EUR. L. 539 (2016), https://ssrn.com/abstract=2807675; Joanna Caytas, Blockchain in the U.S. Regulatory Setting: Evidentiary Use in Vermont, Delaware, and Elsewhere, __ COLUM. SCI. & TECH. L. REV. (2017), https://ssrn.com/abstract=2988363; Jonathan Chiu & Thorsten V. Koeppl, Incentive Compatibility on the Blockchain (May 2018), https://ssrn.com/abstract=3221233; Jonathan Chiu & Thorsten V. Koeppl, Blockchain-Based Settlement for Asset Trading (2018), https://ssrn.com/abstract=3203917; Ryan J. Davies & Erik R. Sirri, The Economics and Regulation of Secondary Trading Markets (July 20, 2017). Presented at Initiating Conference 'New Special Study of the Securities Markets', held at Columbia Law School, 23 & 24 March 2017, https://ssrn.com/abstract=3012536; Simone Fabiano, DLTs: Applications in the Financial Markets (2018), https://ssrn.com/abstract=3202854; Christian P. Fries & Peter Kohl-Landgraf, Smart Derivative Contracts (Detaching Transactions from Counterparty Credit Risk: Specification, Parametrisation, Valuation) (2018), https://ssrn.com/abstract=3163074; Anton Golub, Lidan Grossmass & Ser-Huang Poon, Ultra Short Tenor Yield Curves for High-Frequency Trading and Blockchain Settlement (2018), https://ssrn.com/abstract=3133024; John Michael Grant, Is Bitcoin Money?: Implications for Bitcoin Derivatives Regulation and Security Interest Treatment of Bitcoins Under Article 9 of the Uniform Commercial Code (2014), https://ssrn.com/abstract=2619457; Allan D. Grody, *Rebuilding Financial Industry Infrastructure,* 11 J. RISK MGMT. IN FIN. INSTS. (2018), https://ssrn.com/abstract=3096721; Nikolaus Hautsch, Christoph Scheuch & Stefan Voigt, Limits to Arbitrage in Markets with Stochastic Settlement Latency. CFS Working Paper, No. 616, 2018, https://ssrn.com/abstract=3302159; Robert C. Hockett & Saule T. Omarova, *The Finance Franchise*, 102 CORNELL L. REV. 1143 (2017); https://ssrn.com/abstract=2820176; Jeremy Kress, *Credit Default Swap Clearinghouses and Systematic Risk: Why Centralized Counterparties Must Have*

## II. Smart Cities

The blockchain has been recognized by scholars for its potential to transform urban living.[154]

## JJ. Smart Contracts

Although discussed previously,[155] we list the category of smart contracts again here because of their profound significance to numerous specific uses.[156]

---

*Access to Central Bank Liquidity,* 48 HARV. J. LEGIS. 49 (2011); Larissa Lee, *New Kids on the Blockchain: How Bitcoin's Technology Could Reinvent the Stock Market,* 12(2) HASTINGS BUS. L.J., 81 (2016), https://ssrn.com/abstract=2656501; Mark M. Lennon & Daniel Folkinshteyn, *From Bit Valley to Bitcoin: The NASDAQ Odyssey*, 11(1) GLOBAL J. BUS. RES. 85 (2017), https://ssrn.com/abstract=3025751; Michael Mainelli & Alistair K. L. Milne, The Impact and Potential of Blockchain on the Securities Transaction Lifecycle (2016), https://ssrn.com/abstract=2777404; Katya Malinova & Andreas Park, Market Design with Blockchain Technology (2017), https://ssrn.com/abstract=2785626; Eva Micheler, Explaining the Infrastructure Underpinning Securities Markets – Market Failure and the Role of Technology (2017), https://ssrn.com/abstract=2941643; Eva Micheler & Luke von der Heyde, Holding, Clearing and Settling Securities Through Blockchain Technology Creating an Efficient System by Empowering Asset Owners (2016), https://ssrn.com/abstract=2786972; Philipp Paech, Integrating Global Blockchain Securities Settlement with the Law - Policy Considerations and Draft Principles (2016), https://ssrn.com/abstract=2792639; Gareth Peters & Guy Vishnia, Blockchain Architectures for Electronic Exchange Reporting Requirements: EMIR, Dodd Frank, MiFID I/II, MiFIR, REMIT, Reg NMS and T2S. (2016), https://ssrn.com/abstract=2832604; Andrea Pinna & Wiebe Ruttenberg, Distributed Ledger Technologies in Securities Post-Trading Revolution or Evolution?, ECB Occasional Paper No. 172 (2016), https://ssrn.com/abstract=2770340; Randy Priem, Distributed Ledger Technology for Securities Clearing and Settlement: Benefits, Risks, and Regulatory Implications (2018), https://ssrn.com/abstract=3292815; Angela Walch, *The Bitcoin Blockchain as Financial Market Infrastructure: A Consideration of Operational Risk,* 18 NYU J. LEGIS. & PUB. POL'Y 837 (2015), https://ssrn.com/abstract=2579482.

[154] *See* Marek Banczyk & Jason Potts, City as Neural Platform - Toward New Economics of a City (2018), https://ssrn.com/abstract=3233686.

[155] *See* Section III, *Infra*.

[156] *See* Paul Catchlove, Smart Contracts: A New Era of Contract Use (2017), https://ssrn.com/abstract=3090226; Lin William Cong & Zhiguo He, Blockchain Disruption and Smart Contracts (2018), https://ssrn.com/abstract=2985764; Marco Dell'Erba, Demystifying Technology. Do Smart Contracts Require a New Legal Framework? Regulatory Fragmentation, Self-Regulation, Public Regulation, (2018), https://ssrn.com/abstract=3228445; Helen Eenmaa-Dimitrieva & Maria José Schmidt-Kessen, Regulation Through Code as a Safeguard for Implementing Smart Contracts in No-Trust Environments (2017). EUI Department of Law Research Paper No. 2017/13., https://ssrn.com/abstract=3100181; Guillaume Haeringer & Hanna Halaburda, Bitcoin: A Revolution?, Baruch College Zicklin School of Business Research Paper No. 2018-05-01 (2018), https://ssrn.com/abstract=3133346; Hanna Halaburda, Blockchain Revolution Without the Blockchain, Bank of Canada Staff Analytical Note 2018-5 (2018), https://ssrn.com/abstract=3133313; Richard Holden & Anup Malani, Can Blockchain Solve the Holdup Problem in Contracts?, University of Chicago Coase-Sandor Institute for Law & Economics Research Paper No. 846 (2017), https://ssrn.com/abstract=3093879; Adam J. Kolber, Not-So-Smart Blockchain Contracts and Artificial Responsibility, 21 STAN. TECH. L. REV. 198 (2018),

### KK.  Space Applications

Several scholars have written about the use of blockchain technology for extra-terrestrial applications.[157]

### LL.  State and Municipal Operating Efficiencies

Brookings scholars Kevin C. Desouza, Chen Ye and Kiran Kabtta Somvanshi write, "Now U.S. state governments have recognized the [blochchain] technology's potential for the delivery of public services and are at various stages of implementation. For blockchain to emerge as the technological imperative for public services, states will have to change existing regulations."[158] While the

---

https://ssrn.com/abstract=3186254; Dimitrios Linardatos, Smart Contracts: Some Clarifying Remarks From a German Legal Point of View (2018), https://ssrn.com/abstract=3193588; Jeffrey M. Lipshaw, The Persistence of 'Dumb' Contracts, 2 STAN. J. BLOCKCHAIN L. & POL'Y __ (2019), https://ssrn.com/abstract=3202484; Stephen M. McJohn & Ian McJohn, The Commercial Law of Bitcoin and Blockchain Transactions, UCC L.J. __ (201_), https://ssrn.com/abstract=2874463; Eliza Mik, Smart Contracts: Terminology, Technical Limitations and Real World Complexity (2017), https://ssrn.com/abstract=3038406; Florian Möslein, Legal Boundaries of Blockchain Technologies: Smart Contracts as Self-Help? (2018). A. De Franceschi, R. Schulze, M. Graziadei, O. Pollicino, F. Riente, S. Sica, P. Sirena (eds.), Digital Revolution – New challenges for Law, (2019), https://ssrn.com/abstract=3267852; Florian Möslein, Legal Boundaries of Blockchain Technologies: Smart Contracts as Self-Help?, In A. De Franceschi, R. Schulze, M. Graziadei, O. Pollicino, F. Riente, S. Sica, P. Sirena (eds.), Digital Revolution – New challenges for Law, (2019), https://ssrn.com/abstract=3267852; John M. Newman, Innovation Policy for Cloud-Computing Contracts, Handbook of Research on Digital Transformations (Francisco-Xavier Olleros & Majlinda Zhegu eds., 2016), https://ssrn.com/abstract=2534597; Seth Oranburg & Liya Palagashvili, The Gig Economy, Smart Contracts, and Disruption of Traditional Work Arrangements (2018), https://ssrn.com/abstract=3270867; Maria Letizia Perugini & Paolo Dal Checco, Smart Contracts: A Preliminary Evaluation (2015), https://ssrn.com/abstract=2729548; Max Raskin, The Law and Legality of Smart Contracts, 1 GEO. L. TECH. REV. 304 (2017), https://ssrn.com/abstract=2959166; Jeremy Sklaroff, Smart Contracts and the Cost of Inflexibility, 166 U. PA. L. REV. (2017), https://ssrn.com/abstract=3008899; Melanie Swann & Primavera De Filippi, Toward a Philosophy of Blockchain, In Swan, M. & De Filippi, P., 48(5) Toward a Philosophy of Blockchain in Metaphilosophy (Wiley, 2017), https://ssrn.com/abstract=3097477; Eric Tjong Tjin Tai, Force Majeure and Excuses in Smart Contracts, Tilburg Private Law Working Paper Series No. 10/2018, https://ssrn.com/abstract=3183637; Katrin Tinn, Blockchain and the Future of Optimal Financing Contracts (2017), https://ssrn.com/abstract=3061532; Mark Verstraete, The Stakes of Smart Contracts, 50 LOY. U. CHI. L.J. ___ (forthcoming 2019), https://ssrn.com/abstract=3178393; Hongjiang Zhao & Cephas P.K Coffie, Economic Force of Smart Contracts (2018), https://ssrn.com/abstract=3138063.

[157] *See* Kartik Hegadekatti, Extra-Terrestrial Applications of Blockchains and Cryptocurrencies (2016), https://ssrn.com/abstract=2882763; Mason J. Molesky, Elizabeth A. Cameron, Jerry Jones IV, Michael Esposito, Liran Cohen & Chris Beauregard, *Blockchain Network for Space Object Location Gathering,* IEEE 978-1-5386-7266-2/18.

[158] *See* Kevin C. Desouza, Chen Ye & Kiran Kabtta Somvanshi, *Blockchain and U.S. State Governments: An Initial Assessment*, BROOKINGS (April 17, 2018), https://www.broofkings.edu/blog/techtank/2018/04/17/blockchain-and-u-s-state-governments-an-

activity of many state legislatures thus far has been limited to regulation to clarify existing money transmission laws to reflect cryptocurrency exchanges, Desouza et al., write, "that the vast majority of US states have taken at least some form of regulatory stance concerning cryptocurrencies and blockchain technology."[159] Previously, Trautman has described the Delaware Blockchain Initiative, intended as, "a comprehensive program intended to spur adoption and development of blockchain and smart contract technologies in both public and private sectors."[160] Desouza et al. write:

> [t]hen governor Jack Markell noted that "Smart contracts offer a powerful and innovative way to streamline cumbersome back-office procedures, lower transactional costs for consumers and businesses, and manage and reduce risk," and suggested that the state will "lead the way in promoting blockchain technology and its growing role in digital commerce."[161]

### MM. Tax Computation and Compliance

A number of scholars have explained the use of blockchain technology to facilitate the computation, collection, or other issues surrounding the topic of tax.[162]

---

initial-assessment/; *see also* Joanna Caytas, *Blockchain in the U.S. Regulatory Setting: Evidentiary Use in Vermont, Delaware, and Elsewhere*, COLUM. SCI. & TECH. L. REV. (May 30, 2017), http://stlr.org/2017/05/30/blockchain-in-the-u-s-regulatory-setting-evidentiary-use-in-vermont-delaware-and-elsewhere/.

[159] *Id.*

[160] See Trautman, Bitcoin, Virtual Currencies and the Struggle of Law and Regulation to Keep Pace, 102 MARQ. L. REV. 447, 486 (2018), https://ssrn.com/abstract=3182867, *supra* note 28 at 486, *citing* Desouza, et al.

[161] *Id.*

[162] *See* Richard Thompson Ainsworth & Brendan Magauran, Taxing & Zapping Marijuana: Blockchain Compliance in the Trump Administration Part 1 (2018). B.U. Sch. L., Law and Econ. Res. Paper No. 18-03, https://ssrn.com/abstract=3141203; Richard Thompson Ainsworth & Musaad Alwohaibi, The First Real-Time Blockchain VAT - GCC Solves MTIC Fraud (2017). Boston U. Sch. L., Law and Econ. Res. Paper No. 17-23, https://ssrn.com/abstract=3007753; Rifat Azam & Orly Mazur, Cloudy with a Chance of Taxation, FLA. TAX REV. (Forthcoming), https://ssrn.com/abstract=3290444; David J. Shakow, The Tax Treatment of Tokens: What Does it Betoken?, 156 TAX NOTES 1387 (2017); https://ssrn.com/abstract=3057466; David J. Shakow, *The Tao of the DAO: Taxing an Entity that Lives on a Blockchain*, 160 TAX NOTES 929 (Aug. 13, 2018), https://ssrn.com/abstract=3247155; Manoj Viswanathan, Tax Compliance in a Decentralizing Economy (2017), https://ssrn.com/abstract=3002213.

### NN. Timber Tracking

At least one scholar has focused on the application of blockchain to track timber production.[163]

### VII. CONCLUSION

The blockchain is a new and potentially disruptive technology. We have looked at a number of promising use cases. Our goal has been to provide a highly readable description of what blockchain entails; how it works; and what blockchain applications show promise beyond virtual currencies. We envision entrepreneurs coming together with those trained in computer science and cryptography to conduct interdisciplinary research and product development. The development of realistic blockchain-based systems is underway as the blockchain community gains traction.

---

[163] *See* Boris Düdder & Omri Ross, Timber Tracking: Reducing Complexity of Due Diligence by Using Blockchain Technology (2017), https://ssrn.com/abstract=3015219.